\pdfoutput=1

\documentclass[twocolumn]{svjour3}          % twocolumn
\smartqed  % flush right qed marks, e.g. at end of proof

\usepackage[longnamesfirst]{natbib}
\usepackage{amsmath, amsbsy, epsfig, epsf, psfrag, booktabs, graphicx, amssymb, rotating, placeins, subfig}
\usepackage[table,x11names]{xcolor}
\usepackage[tablesfirst, nolists]{endfloat}

\usepackage{alltt}

\usepackage{authblk}

\newcommand{\bm}[1]{{\boldsymbol {#1}}}
\newcommand{\bZ}{{\bm Z}}

\newcommand{\dif}{\mathrm{d}}%

\makeatletter
\def\old@comma{,}
\catcode`\,=13
\def,{%
  \ifmmode%
    \old@comma\discretionary{}{}{}%
  \else%
    \old@comma%
  \fi%
}
\makeatother

\newtheorem{thm}{Theorem}

\graphicspath{{./figs/}}
\DeclareGraphicsExtensions{.eps,.ps,.pdf}

\begin{document}

\title{Automated Selection of $r$ for the $r$ Largest Order Statistics
  Approach with Adjustment for Sequential Testing
\thanks{This research was partially supported by an NSF grant 
  (DMS 1521730), a University of Connecticut Research Excellence
  Program grant, and a contract with Environment and Climate Change Canada.}
  % % Grants or other notes
  % % about the article that should go on the front page should be
  % % placed here. General acknowledgments should be placed at the end of
  % % the article.}
}

\author{Brian Bader \and Jun Yan \and Xuebin Zhang}

\institute{Brian Bader \at
           University of Connecticut \\
           \email{brian.bader@uconn.edu}
           \and Jun Yan      \at
           University of Connecticut \\
           \email{jun.yan@uconn.edu}
      	   \and Xuebin Zhang \at
           Environment and Climate Change Canada \\
           \email{xuebin.zhang@canada.ca}
}

%\date{Received: date / Accepted: date}
% The correct dates will be entered by the editor

\maketitle

\begin{abstract}
The $r$ largest order statistics approach is widely used in 
extreme value analysis because it may use more information 
from the data than just the block maxima. 
In practice, the choice of $r$ is critical. 
If $r$ is too large, bias can occur; if too small, 
the variance of the estimator can be high. 
The limiting distribution of the $r$ largest order statistics, 
denoted by GEV$_r$, extends that of the block maxima. 
Two specification tests are proposed to select $r$ sequentially. 
The first is a score test for the GEV$_r$ distribution. 
Due to the special characteristics of the GEV$_r$ distribution, 
the classical chi-square asymptotics cannot be used. 
The simplest approach is to use the parametric bootstrap, which is 
straightforward to implement but computationally expensive. 
An alternative fast weighted bootstrap or multiplier 
procedure is developed for computational efficiency. 
The second test uses the difference in estimated entropy between 
the GEV$_r$ and GEV$_{r-1}$ models, applied to the $r$ largest order 
statistics and the $r-1$ largest order statistics, respectively. 
The asymptotic distribution of the difference statistic is derived.
In a large scale simulation study, both tests held their size and 
had substantial power to detect various misspecification schemes. 
A new approach to address the issue of multiple, sequential 
hypotheses testing is adapted to this setting to control the 
false discovery rate or familywise error rate.
The utility of the procedures is demonstrated with extreme sea level 
and precipitation data.

%% IMPORTANT?? Comments from the template file:
%Include PACS and mathematical subject classification numbers as needed.

\keywords{entropy \and generalized extreme value \and goodness-of-fit \and 
multiplier bootstrap \and score test \and sequential testing}

\end{abstract}

\section{Introduction}
\label{sect:intr}

% r-LOS method
The largest order statistics approach is an extension of the block 
maxima approach that is often used in extreme value modeling.
The focus of this paper is \citep[p.28--29]{smith1986extreme}:
``Suppose we are given, not just the maximum value for each year,
but the largest ten (say) values. How might we use this data to obtain
better estimates than could be made just with annual maxima?''
The $r$ largest order statistics approach may use more information 
than just the block maxima in extreme value analysis, and is widely
used in practice when such data are available for each block.
The approach is based on the limiting distribution of the $r$ 
largest order statistics which extends the generalized extreme 
value (GEV) distribution \citep[e.g.,][]{weissman1978estimation}. 
This distribution, denoted by GEV$_r$, has the same parameters
as the GEV distribution, which makes it useful to estimate the GEV 
parameters when the $r$ largest values are available for each block.
The approach was investigated by \citet{smith1986extreme} for the 
limiting joint Gumbel distribution and extended to the more general 
limiting joint GEV$_r$ distribution by \citet{tawn1988extreme}.
Because of the potential gain in efficiency relative to the block
maxima only, the method has found many applications such as
corrosion engineering \citep[e.g.,][]{scarf1996estimation}, 
hydrology \citep[e.g.,][]{dupuis1997extreme},
coastal engineering \citep[e.g.,][]{guedes2004application},
and wind engineering \citep[e.g.,][]{an2007r}.

% selection of r is and is not
In practice, the choice of $r$ is a critical issue in extreme 
value analysis with the $r$ largest order statistics approach. 
In general $r$ needs to be small relative to the block size
$B$ (not the number of blocks $n$) because as $r$ increases,
the rate of convergence to the limiting joint distribution 
decreases sharply \citep{smith1986extreme}. 
There is a trade-off between the validity of the limiting result
and the amount of information required for good estimation.
If $r$ is too large, bias can occur; if too small, 
the variance of the estimator can be high.
Finding the optimal $r$ should lead to more efficient
estimates of the GEV parameters without introducing bias. 
A much related but different problem is the selection of 
threshold or fraction of a sample extreme value analysis
\citep[see][for a review]{scarrott2012review}.
Our focus here is the selection of $r$ for situations where 
a number of largest values are available each of $n$ blocks.
In contrast, the methods for threshold or fraction selection 
reviewed in \citet{scarrott2012review} deal with
a single block ($n = 1$) of a large size $B$.

% existing work
The selection of $r$ has not been as actively researched as 
the threshold selection problem in the one sample case. 
\citet{smith1986extreme} and \citet{tawn1988extreme} used 
probability (also known as PP) plots for the marginal distribution
of the $r$th order statistic to assess its goodness of fit. 
The probability plot provides a visual diagnosis, but different viewers 
may reach different conclusions in the absence of a p-value.
Further, the probability plot is only checking the marginal 
distribution for a specific $r$ as opposed to the joint distribution.
\citet{tawn1988extreme} suggested an alternative test of fit
using a spacings results in \citet{weissman1978estimation}.
Let $D_{n:i}$ be the spacing between the $i$th and $(i+1)$th
largest value in a sample of size $B$ from a distribution 
in the domain of attraction of the Gumbel distribution. 
Then $\{iD_i: i = 1, \ldots, r-1\}$ is approximately a set
of independent and identically distributed exponential random 
variables as $B \to \infty$. 
The connections among the three limiting forms of the GEV 
distribution \citep[e.g.,][p.123]{embrechts1997modelling}
can be used to transform from the Fr\'echet and the Weibull
distribution to the Gumbel distribution.
Testing the exponentiality of the spacings on the Gumbel
scale provides an approximate diagnosis of the joint distribution
of the $r$ largest order statistics when $B$ is large.
A limitation of this method, however, is that prior knowledge
of the domain of attraction of the distribution is needed. 
Lastly, \citet{dupuis1997extreme} proposed a robust 
estimation method, where the weights can be used to 
detect inconsistencies with the GEV$_r$ distribution
and assess the fit of the data to the joint Gumbel model.
The method can be extended to general GEV$_r$ distributions
but the construction of the estimating equations is computing
intensive with Monte Carlo integrations.

% preview of the contribution 
In this paper, two specification tests are proposed to select 
$r$ through a sequence of hypothesis testing.
The first is the score test \citep[e.g.,][]{rao2005score}, but 
because of the nonstandard setting of the GEV$_r$ distribution, 
the usual $\chi^2$ asymptotic distribution is invalid. 
A parametric bootstrap can be used to assess the significance
of the observed statistic, but is computationally demanding. 
A fast, large sample alternative to parametric bootstrap based on 
the multiplier approach \citep{kojadinovic2012goodness} is developed. 
The second test uses the difference in estimated entropy between
the GEV$_r$ and GEV$_{r-1}$ models, applied to the $r$ largest order 
statistics and the $r-1$ largest order statistics, respectively. 
The asymptotic distribution is derived with the central limit theorem. 
Both tests are intuitive to understand, easy to implement, and have 
substantial power as shown in the simulation studies.
Each of the two tests is carried out to test the adequacy 
of the GEV$_r$ model for a sequence of $r$ values. 
The very recently developed stopping rules for ordered
hypotheses in \citet{g2015sequential} are adapted to control 
the false discovery rate (FDR), the expected proportion 
of incorrectly rejected null hypotheses among all rejections,
or familywise error rate (FWER), the probability of at least one 
type~I error in the whole family of tests.
All the methods are available in an R package \texttt{eva}
\citep{Rpkg:eva}.

% outline
The rest of the article is organized as follows. 
The problem is set up in Section~\ref{s:setup} with the GEV$_r$ 
distribution, observed data, and the hypothesis to be tested.
The score test is proposed in Section~\ref{s:score} with two 
implementations: parametric bootstrap and multiplier bootstrap. 
The entropy difference (ED) test is proposed and the asymptotic
distribution of the testing statistic is derived in Section~\ref{s:ed}.
A large scale simulation study on the empirical size and power of 
the tests are reported in Section~\ref{s:sim}. 
In Section~\ref{s:seq}, the multiple, sequential testing problem
is addressed by adapting recent developments on this application.
The tests are applied to sea level and precipitation datasets 
in Section~\ref{s:app}. A discussion concludes in Section~\ref{s:disc}. 
The Appendix contains the details of random number generation 
from the GEV$_r$ distribution and a sketch of the proof of the
asymptotic distribution of the ED test statistic.

\section{Model and Data Setup}
\label{s:setup}

The limit joint distribution of the $r$ largest order statistics of 
a random sample of size $B$ as $B \to \infty$ is the GEV$_r$ 
distribution with density function~\citep{weissman1978estimation}
\begin{equation}
\label{eq:gevr}
\begin{split}
& f_r (x_1, x_2, ..., x_r | \mu, \sigma, \xi)\\
= & \sigma^{-r} \exp\Big\{-(1+\xi z_r)^{-\frac{1}{\xi}}  - \left(\frac{1}{\xi}+1\right)\sum_{j=1}^{r}\log(1+\xi z_j)\Big\}
\end{split}
\end{equation}
for some location parameter $\mu$, scale parameter $\sigma > 0$
and shape parameter $\xi$, 
where $x_1 >  \cdots> x_r$, $z_j = (x_j - \mu) / \sigma$, 
and $ 1 + \xi z_j > 0 $ for $j=1, \ldots, r$.
When $r = 1$,  this distribution is exactly the GEV distribution.
The parameters $\theta = (\mu, \sigma, \xi)^{\top}$ remain the 
same for $j = 1, \ldots, r$, $r \ll B$, but the convergence rate
to the limit distribution reduces sharply as $r$ increases.
The conditional distribution of the $r$th component given the top 
$r - 1$ variables in~\eqref{eq:gevr} is the GEV distribution right
truncated by $x_{r-1}$, which facilitates simulation from the GEV$_r$
distribution; see Appendix~\ref{s:gevrsim}.

The $r$~largest order statistics approach is an extension of the 
block maxima approach in extreme value analysis when a number
of largest order statistics are available for each one of a collection 
of independent blocks \citep{smith1986extreme,tawn1988extreme}.
Specifically, let $(x_{i1}, \ldots, x_{ir})$ be the observed $r$~largest 
order statistics from block $i$ for $i = 1, \ldots, n$.
Assuming independence across blocks, the GEV$_r$ distribution is 
used in place of the GEV distribution in the block maxima approach
to make likelihood-based inference about $\theta$.
Let $l_i^{(r)} (\theta) = l^{(r)} (x_{i1}, \ldots, x_{ir} | \theta)$,
where
\begin{equation}
\label{eq:ll}
\begin{split}
& l^{(r)} (x_1, \ldots, x_r | \theta) \\
=& -r\log{\sigma} - (1+\xi z_r)^{-\frac{1}{\xi}}  - \left(\frac{1}{\xi}+1\right)\sum_{j=1}^{r}\log(1+\xi z_j)
\end{split}
\end{equation}
is the contribution to the log-likelihood from a single block
$(x_1, \ldots, x_r)$.
The maximum likelihood estimator (MLE) of $\theta$ using the
$r$ largest order statistics is
$\hat\theta_n^{(r)} = \arg\max \sum_{i=1}^n l_i^{(r)} (\theta)$.

Model checking is a necessary part of statistical analysis.
The rationale of choosing a larger value of $r$ is to use 
as much information as possible, but not set $r$ too
high so that the GEV$_r$ approximation becomes poor
due to the decrease in convergence rate.
Therefore, it is critical to test the goodness-of-fit of 
the GEV$_r$ distribution with a sequence of null hypotheses
\begin{center}
$H_0^{(r)}$: the GEV$_r$ distribution fits the sample of the
$r$ largest order statistics well
\end{center}
for $r=1, \ldots, R$, where $R$ is the maximum, 
predetermined number of top order statistics to test. 
Two test procedures for $H_0^{(r)}$ are developed for 
a fixed $r$ first to help choose $r \geq 1$ such that 
the GEV$_r$ model still adequately describes the data.
The sequential testing process and the multiple testing
issue are investigated in Section~\ref{s:seq}.

\section{Score Test}
\label{s:score}

A score statistic for testing goodness-of-fit hypothesis
$H_0^{(r)}$ is constructed in the usual way with the score 
function and the Fisher information matrix
\citep[e.g.,][]{rao2005score}.
For ease of notation, the superscript $(r)$ is dropped.
Define the score function
\[
S(\theta) = \sum_{i=1}^n S_i(\theta) = 
\sum_{i=1}^n \partial l_i(\theta) / \partial \theta
\]
and Fisher information matrix $I(\theta)$, 
which have been derived in \citet{tawn1988extreme}. 
The behaviour of the maximum likelihood estimator is 
the same as that derived for the block maxima approach 
\citep{smith1985maximum,tawn1988extreme}, which
requires $\xi > -0.5$.
The score statistic is
\begin{displaymath}
V_n = \frac{1}{n} S^{\top}(\hat\theta_n) I^{-1}(\hat\theta_n) S(\hat\theta_n).
\end{displaymath}

\begin{figure*}[tbp]
    \centering
    \includegraphics[width=\textwidth]{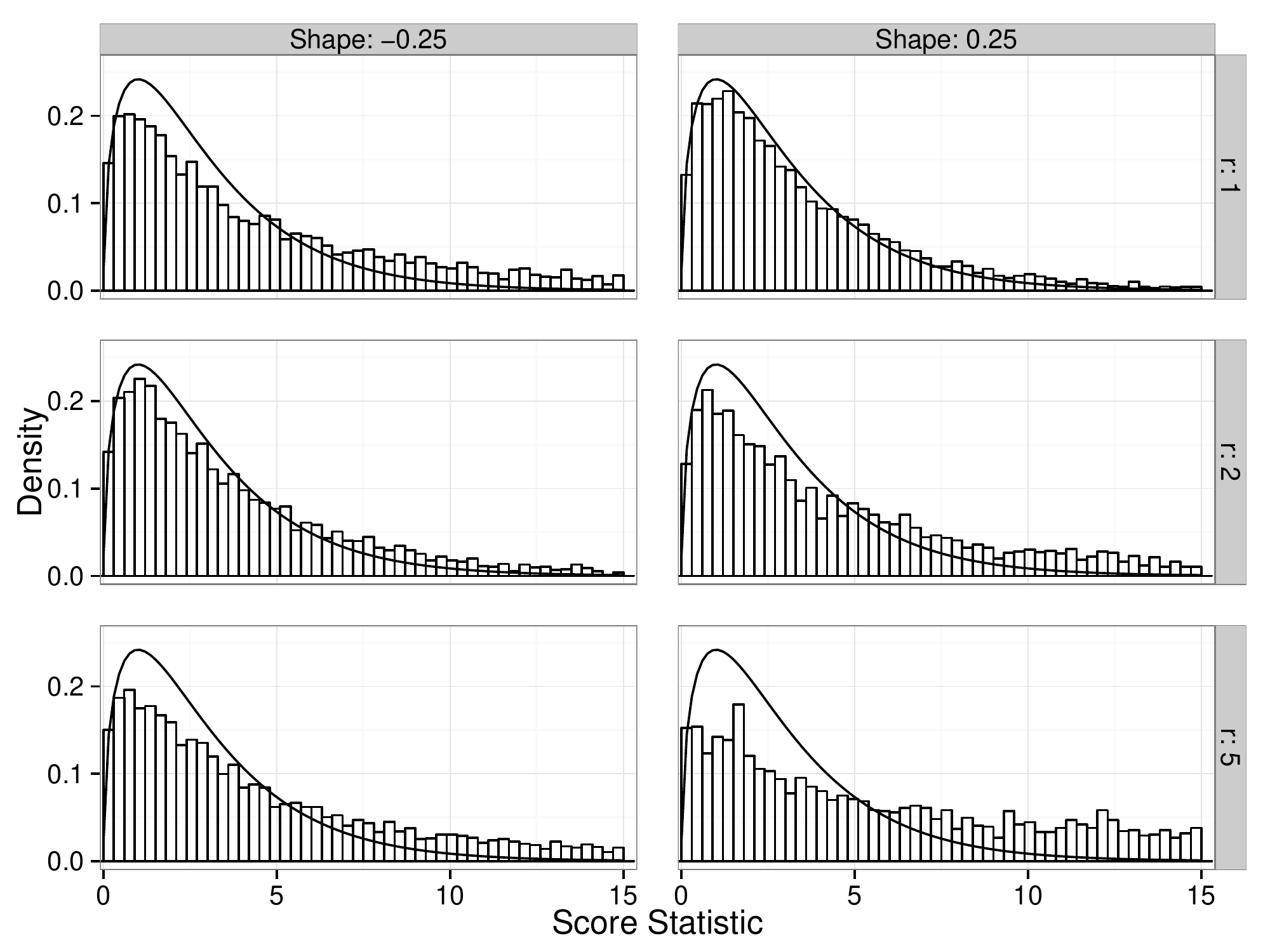}
    \caption{Comparisons of the empirical distribution based on 5000
      replicates of the score test statistic and the $\chi^2(3)$
      distribution (red solid curve). The number of blocks used is $n =5000$ 
      with parameters $\mu=0$, $\sigma=1$, and 
      $\xi \in (-0.25, 0.25)$.
    }
    \label{fig:AsymChiSq}
\end{figure*}

Under standard regularity conditions, $V_n$ would asymptotically
follow a $\chi^2$ distribution with 3 degrees of freedom. 
The GEV$_r$ distribution, however, violates the regularity conditions 
for the score test \citep[e.g.,][pp. 516-517]{casella2002statistical},
as its support depends on the parameter values unless $\xi = 0$.
For illustration, Figure~\ref{fig:AsymChiSq} presents a visual 
comparison of the empirical distribution of $V_n$ with $n = 5000$
from 5000 replicates, overlaid with the $\chi^2(3)$ distribution,
for $\xi \in \{-0.25, 0.25\}$ and $r \in \{1, 2, 5\}$.
The sampling distribution of $V_n$ appears to be much heavier tailed
than $\chi^2(3)$, and the mismatch increases as $r$ increases as a
result of the reduced convergence rate.

Although the regularity conditions do not hold, the score 
statistic still provides a measure of goodness-of-fit since 
it is a quadratic form of the score, which has expectation 
zero under the null hypothesis.
Extremely large values of $V_n$ relative to its sampling
distribution would suggest lack of fit, and, hence, 
possible misspecification of $H_0^{(r)}$.
So the key to applying the score test is to get an 
approximation of the sampling distribution of $V_n$.
Two approaches for the approximation are proposed.

\subsection{Parametric Bootstrap}
\label{s:pb}

The first solution is parametric bootstrap.
For hypothesis $H_0^{(r)}$, the test procedure goes as follows:
\begin{enumerate}
\item
Compute $\hat\theta_n$ under $H_0$ with the observed data.
\item
Compute the testing statistic $V_n$.
\item
For every $k \in  \{1, ..., L \}$ with a large number $L$, repeat:
  \begin{enumerate}
  \item
  Generate a bootstrap sample of size $n$ for the $r$ largest statistics 
  from GEV$_r$ with parameter vector $\hat\theta_n$.
  \item
  Compute the $\hat\theta_n^{(k)}$ under $H_0$ with the bootstrap sample.
  \item
  Compute the score test statistic $V_n^{(k)}$.
  \end{enumerate}
\item
Return an approximate p-value of $V_n$ as \\ %break this line manually
$L^{-1} \sum_{k=1}^{L} 1(V_n^{(k)} > V_n)$.
\end{enumerate}

Straightforward as it is, the parametric bootstrap approach
involves sampling from the null distribution and computing the MLE 
for each bootstrap sample, which can be very computationally expensive.
This is especially true as the sample size $n$ and/or the number of
order statistics $r$ included in the model increases.

\subsection{Multiplier Bootstrap}
\label{s:mb}

Multiplier bootstrap is a fast, large sample alternative to parametric 
bootstrap in goodness-of-fit testing \citep[e.g.,][]{kojadinovic2012goodness}.
The idea is to approximate the asymptotic distribution of 
$n^{-1/2} I^{-1/2}(\theta) S(\theta)$ using its asymptotic representation
\[
n^{-1/2} I^{-1/2}(\theta) S(\theta) = 
\frac{1}{\sqrt{n}}\sum_{i=1}^n \phi_i(\theta),
\]
where $\phi_i(\theta) =  I^{-1/2} (\theta) S_i(\theta)$.
Its asymptotic distribution is the same as the asymptotic distribution of
\[
W_n(\bZ, \theta) = \frac{1}{\sqrt{n}} \sum_{i=1}^n (Z_i  - \bar Z)\phi_i(\theta),
\]
conditioning on the observed data, 
where ${\bZ} = (Z_1, ..., Z_n)$ is a set of independent and identically
distributed multipliers (independent of the data), with expectation 0 and
variance 1, and $\bar{Z} = \frac{1}{n} \sum_{i=1}^{n} Z_i$. 
The multipliers must satisfy 
$\int_0^\infty \{\Pr(|Z_1| > x)\}^{\frac{1}{2}} \dif x < \infty $. 
An example of a possible multiplier distribution is $N(0, 1)$.

The multiplier bootstrap test procedure is summarized as follows:
\begin{enumerate}
\item
Compute $\hat\theta_n$ under $H_0$ with the observed data.
\item
Compute the testing statistic $V_n$.
\item
For every $k \in  \{1, ..., L \}$ with a large number $L$, repeat:
  \begin{enumerate}
  \item
    Generate $\bZ^{(k)} = (Z_1^{(k)}, \ldots, Z_n^{(k)})$ from $N(0, 1)$.
  \item
    Compute a realization from the approximate distribution of 
    $W_n(\bZ, \theta)$ with $W_n(\bZ^{(k)}, \hat\theta_n)$.
  \item 
    Compute $V_n^{(k)}(\hat\theta_n) = W_n^{\top}(\bZ^{(k)}, \hat\theta_n) W_n(\bZ^{(k)}, \hat\theta_n)$.
  \end{enumerate}
\item
Return an approximate p-value of $V_n$ as \\ %manual break
$L^{-1} \sum_{k=1}^{L} 1(V_n^{(k)} > V_n)$.
\end{enumerate}

This multiplier bootstrap procedure is much faster than parametric
bootstrap procedure because, for each sample, it only
needs to generate $\bZ$ and compute $W_n(\bZ, \hat\theta_n)$.
The MLE only needs to be obtained once from the observed data.

\section{Entropy Difference Test}
\label{s:ed}

Another specification test for the GEV$_r$ model is derived based 
on the  difference in entropy for the GEV$_r$ and GEV$_{r-1}$ models. 
The entropy for a continuous random variable with density $f$ is
\citep[e.g.,][]{singh2013entropy}
\begin{displaymath}
E[-\ln  f(y)] = - \int_{-\infty}^{\infty}  f(y) \log f(y)   \dif y.
\end{displaymath}
It is essentially the expectation of negative log-likelihood.
The expectation can be approximated with the sample average of
the contribution to the log-likelihood from the observed data,
or simply the log-likelihood scaled by the sample size $n$.
Assuming that the $r - 1$ top order statistics fit the GEV$_{r-1}$
distribution, the difference in the log-likelihood between GEV$_{r-1}$ 
and GEV$_r$ provides a measure of deviation from $H_0^{(r)}$.
Its asymptotic distribution can be derived.
Large deviation from the expected difference under $H_0^{(r)}$
suggests a possible misspecification of $H_0^{(r)}$.

From the log-likelihood contribution in~\eqref{eq:ll}, 
the difference in log-likelihood for the $i$th block,
$Y_{ir} (\theta) = l_i^{(r)}  - l_i^{(r-1)}$, is 
\begin{align}
\label{eq:dll}
Y_{ir}(\theta) = &-\log{\sigma} - (1+\xi z_{ir})^{-\frac{1}{\xi}}
                   + (1+\xi z_{i{r-1}})^{-\frac{1}{\xi}}    \notag \\
& - \bigl(\frac{1}{\xi}+1 \bigr)\log(1+\xi z_{ir}).
\end{align}
Let $\bar{Y_r} = \frac{1}{n} \sum_{i=1}^{n} Y_{ir}$ and 
$S_{Y_r}^2 = \sum_{i=1}^{n} (Y_{ir} - \bar{Y_r})^2 / (n - 1)$
be the sample mean and sample variance, respectively.
Consider a standardized version of $\bar Y_r$ as
\begin{equation}
  \label{eq:ed}
  T^{(r)}_n (\theta) = \sqrt{n}(\bar Y_r - \eta_r) / S_{Y_r},
\end{equation}
where $\eta_r = -\log{\sigma} - 1 + (1+\xi)\psi(r)$, and
$\psi(x) = \dif \log \Gamma(x) / \dif x$ is the digamma function.
The asymptotic distribution of $T^{(r)}_n$ is summarized by 
Theorem~\ref{thm:ed} whose proof is relegated to Appendix~\ref{s:tn}.

\begin{thm}
\label{thm:ed}
Let $T_n^{(r)}(\theta)$ be the quantity computed based on a random 
sample of size $n$ from the GEV$_r$ distribution with parameters $\theta$ 
and assume that $H_0^{(r-1)}$ is true. Then $T_n^{(r)}$ converges in 
distribution to $N(0, 1)$ as $n \to \infty$.
\end{thm}

Note that in Theorem~\ref{thm:ed}, $T_n^{(r)}$ is computed 
from a random sample of size $n$ from a GEV$_r$ distribution. 
If the random sample were from a distribution in the domain of
attraction of a GEV distribution, the quality of the approximation
of the GEV$_r$ distribution to the $r$ largest order statistics
depends on the size of each block $B \to \infty$ with $r \ll B$.
The block size $B$ is not to be confused with the sample size $n$. 
Assuming $\xi > -0.5$, the proposed ED statistic for $H_0^{(r)}$ 
is $T_n^{(r)} (\hat\theta_n)$, where $\hat\theta_n$ is the MLE of 
$\theta$ with the $r$ largest order statistics for the 
GEV$_{r}$ distribution. Since $\hat\theta_n$ is consistent
for $\theta$ with $\xi > -0.5$, $T_n^{(r)}(\hat\theta_n)$ has the same
limiting distribution as $T_n^{(r)}(\theta)$ under $H_0^{(r)}$.

To assess the convergence of $T_n^{(r)}(\hat\theta_n)$ to $N(0, 1)$,
1000 GEV$_r$ replicates were simulated under configurations of 
$r \in \{2, 5, 10\}$, $\xi \in \{-0.25, 0, 0.25\}$, and 
$n \in \{50, 100\}$. Their quantiles are compared with those of 
$N(0, 1)$ via quantile-quantile plots (not presented). 
It appears that a larger sample size is needed for the normal
approximation to be good for larger $r$ and negative $\xi$.
This is expected because larger $r$ means higher dimension 
of the data, and because the MLE only exists for $\xi > -0.5$
\citep{smith1985maximum}. For $r$ less than 5 and $\xi \ge 0$, 
the normal approximation is quite good; it appears satisfactory 
for sample size as small as 50. For $r$ up to 10, sample size 
100 seems to be sufficient.

\section{Simulation Results}
\label{s:sim}

\subsection{Size}
\label{ss:size}
The empirical sizes of the tests are investigated first.
For the score test, the parametric bootstrap version and 
the multiplier bootstrap version are equivalent asymptotically,
but may behave differently for finite samples.
It is of interest to know how large a sample size is needed 
for the two versions of the score test to hold their levels.
Random samples of size $n$ were generated from the 
GEV$_r$ distribution with $r \in \{1, 2, 3, 4, 5, 10\}$, 
$\mu = 0$, $\sigma = 1$, and $\xi \in \{-0.25, 0, 0.25\}$.
All three parameters $(\mu, \sigma, \xi)$ were estimated.

When the sample size is small, there can be numerical difficulty 
in obtaining the MLE. 
For the multiplier bootstrap score and ED test, the MLE only 
needs to obtained once, for the dataset being tested. However, in 
addition, the parametric bootstrap score test must obtain 
a new sample and obtain the MLE for each bootstrap replicate. 
To assess the severity of this issue, 10,000 datasets were 
simulated for $\xi \in \{-0.25, 0, 0.25\}$, $r \in \{1,2,3,4,5,10\}$, 
$n \in \{25, 50\}$, and the MLE was attempted for each dataset. 
Failure never occurred for $\xi \geq 0$. With $\xi = -0.25$ and 
sample size 25, the highest failure rate of 0.69\% occurred for 
$r=10$. When the sample size is 50, failures only occurred 
when $r=10$, at a rate of 0.04\%.

For the parametric bootstrap score test with sample size
$n \in \{25, 50, 100\}$,
Table~\ref{tab:pbsize} summarizes the empirical size of the 
test at nominal levels 1\%, 5\%, and 10\% obtained from 1000 
replicates, each carried out with bootstrap sample size $L = 1000$. 
Included only are the cases that converged successfully. Otherwise, 
the results show that the agreement between the empirical levels 
and the nominal level is quite good for samples as small as 25, 
which may appear in practice when long record data is not available.

\begin{table}[tbp]
  \centering
  \caption{Empirical size (in \%) for the parametric bootstrap score
    test under the null distribution GEV$_r$, with $\mu=0$ and
    $\sigma=1$ based on 1000 samples, each with bootstrap sample 
    size $L = 1000$.}
    \begin{tabular}{cc rrr rrr rrr}
      \toprule
      Sample Size & $r$ & \multicolumn{3}{c}{25} & \multicolumn{3}{c}{50} & \multicolumn{3}{c}{100} \\
      \cmidrule(lr){3-5}\cmidrule(lr){6-8}\cmidrule(lr){9-11}
      Nominal Size & & 1.0 & 5.0 & 10.0  & 1.0 & 5.0 & 10.0  & 1.0 & 5.0 & 10.0 \\
      \midrule
    $\xi=-0.25$ & 1     & 0.4   & 2.8   & 6.0   & 1.1   & 4.8   & 9.3   & 0.6   & 4.1   & 8.0 \\
                & 2     & 0.1   & 2.6   & 6.0   & 0.8   & 3.4   & 6.5   & 0.6   & 3.6   & 8.1 \\
                & 3     & 0.3   & 2.5   & 5.0   & 0.8   & 4.3   & 7.7   & 1.1   & 4.8   & 8.1 \\
                & 4     & 0.3   & 1.8   & 5.4   & 0.6   & 3.1   & 6.9   & 1.1   & 5.1   & 8.8 \\
                & 5     & 0.4   & 2.4   & 6.7   & 0.4   & 3.3   & 8.3   & 0.6   & 3.1   & 6.5 \\
                & 10    & 2.7   & 5.3   & 8.7   & 0.5   & 3.9   & 8.4   & 0.7   & 4.2   & 7.6 \\[6pt]
    $\xi=0$     & 1     & 1.3   & 5.2   & 8.9   & 1.6   & 5.3   & 9.0   & 0.8   & 4.7   & 9.3 \\
                & 2     & 1.4   & 5.1   & 9.4   & 2.0   & 4.9   & 10.0  & 1.0   & 4.3   & 9.9 \\
                & 3     & 1.7   & 6.2   & 10.9  & 2.1   & 6.0   & 10.2  & 0.8   & 4.9   & 9.8 \\
                & 4     & 1.5   & 4.5   & 8.5   & 1.3   & 6.0   & 10.2  & 1.0   & 4.4   & 9.8 \\
        	    & 5     & 1.6   & 5.8   & 10.4  & 2.4   & 6.2   & 9.9   & 1.2   & 5.0   & 9.7 \\
        	    & 10    & 1.5   & 4.0   & 7.3   & 1.5   & 4.3   & 8.9   & 0.7   & 4.6   & 8.2 \\[6pt]
    $\xi=0.25$  & 1     & 1.7   & 4.5   & 9.7   & 2.6   & 7.1   & 11.5  & 1.1   & 4.6   & 9.1 \\
       	        & 2     & 1.8   & 5.1   & 8.7   & 1.8   & 4.4   & 8.5   & 0.5   & 2.9   & 7.5 \\
       		    & 3     & 1.5   & 4.4   & 9.4   & 1.5   & 3.7   & 8.1   & 1.0   & 4.2   & 9.4 \\
       		    & 4     & 1.2   & 3.3   & 8.1   & 1.1   & 4.6   & 9.7   & 1.1   & 4.3   & 9.6 \\
       		    & 5     & 1.7   & 4.4   & 9.4   & 1.1   & 4.2   & 8.6   & 0.6   & 4.8   & 9.6 \\
       		    & 10    & 1.1   & 4.6   & 8.3   & 1.5   & 6.1   & 10.7  & 1.0   & 3.9   & 8.5 \\
    \bottomrule
    \end{tabular}
  \label{tab:pbsize}
\end{table}

For the multiplier bootstrap score test, the results for
sample sizes $n \in \{25, 50, 100, 200, 300, 400\}$ 
are summarized in Table~\ref{tab:multsize}. 
When the sample size is less than 100, it appears that there
is a large discrepancy between the empirical and nominal level.
For $\xi \in \{0, 0.25\}$, there is reasonable agreement between the 
empirical level and the nominal levels for sample size at least 100.
For $\xi = - 0.25$ and sample size at least 100, the agreement
is good except for $r=1$, in which case, the empirical level is 
noticeably larger than the nominal level. 
This may be due to different rates of convergence for
various $\xi$ values as $\xi$ moves away from $-0.5$.
It is also interesting to note that, everything else being 
held, the agreement becomes better as $r$ increases.
This may be explained by the more information provided by larger $r$
for the same sample size $n$, as can be seen directly in the fisher
information matrix \citep[pp. 247--249]{tawn1988extreme}. 
For the most difficult case with $\xi = -0.25$ and $r = 1$,
the agreement gets better as sample size increases and becomes 
acceptable when sample size was 1000 (not reported).

\begin{table}[tbp]
    \centering
    \caption{Empirical size (in \%) for multiplier bootstrap score
      test under the null distribution GEV$_r$, with $\mu=0$ and
      $\sigma=1$. 1000 samples, each with bootstrap sample size
      $L = 1000$ were used. Although not shown, the empirical 
      size for $r=1$ and $\xi=-0.25$ becomes acceptable when 
      sample size is 1000.}
    \begin{tabular}{cc rrr rrr rrr}
      \toprule
      Sample Size & $r$ & \multicolumn{3}{c}{25} & \multicolumn{3}{c}{50} & \multicolumn{3}{c}{100} \\
      \cmidrule(lr){3-5}\cmidrule(lr){6-8}\cmidrule(lr){9-11}
      Nominal Size & & 1.0 & 5.0 & 10.0  & 1.0 & 5.0 & 10.0  & 1.0 & 5.0 & 10.0  \\
      \midrule
    $\xi=-0.25$ 	 & 1     & 7.0   & 13.4  & 18.9  & 6.3   & 13.8  & 19.6  & 5.4   & 11.4  & 16.3 \\
                     & 2     & 2.0   & 6.9   & 13.4  & 1.3   & 6.4   & 12.4  & 1.6   & 6.9   & 13.6 \\
                     & 3     & 2.1   & 5.8   & 11.7  & 1.1   & 5.9   & 11.1  & 1.1   & 5.0   & 10.8 \\
                     & 4     & 3.3   & 7.2   & 12.3  & 1.1   & 4.9   & 10.8  & 1.0   & 5.2   & 11.9 \\
                     & 5     & 3.6   & 9.0   & 14.0  & 2.3   & 6.8   & 11.2  & 1.1   & 6.2   & 10.6 \\
                     & 10    & 2.0   & 7.0   & 10.3  & 2.6   & 7.4   & 12.8  & 2.1   & 6.4   & 10.1 \\
    $\xi=0$  		 & 1     & 3.3   & 8.4   & 15.3  & 2.2   & 7.0   & 12.5  & 1.1   & 4.6   & 9.2 \\
                     & 2     & 2.8   & 8.7   & 14.4  & 1.8   & 7.5   & 13.0  & 0.9   & 5.7   & 10.3 \\
                     & 3     & 6.1   & 12.1  & 16.5  & 3.0   & 7.2   & 12.2  & 1.5   & 6.0   & 10.4 \\
                     & 4     & 5.1   & 10.4  & 14.5  & 3.6   & 10.1  & 14.9  & 1.0   & 5.6   & 10.3 \\
                     & 5     & 4.2   & 9.0   & 14.5  & 2.2   & 8.2   & 12.5  & 1.7   & 6.5   & 12.0 \\
                     & 10    & 3.1   & 9.2   & 14.4  & 2.4   & 6.4   & 9.8   & 0.6   & 4.6   & 9.0 \\
    $\xi=0.25$ 		 & 1     & 1.8   & 6.7   & 13.7  & 1.3   & 4.7   & 10.4  & 0.8   & 4.4   & 11.5 \\
                     & 2     & 5.7   & 12.7  & 17.1  & 4.7   & 9.9   & 14.9  & 3.5   & 7.4   & 11.6 \\
                     & 3     & 7.1   & 12.2  & 16.5  & 5.3   & 9.4   & 14.8  & 4.2   & 8.4   & 12.5 \\
                     & 4     & 5.4   & 9.8   & 16.8  & 3.7   & 9.0   & 13.4  & 2.6   & 6.0   & 11.4 \\
                     & 5     & 4.4   & 10.1  & 15.8  & 3.5   & 8.2   & 13.6  & 2.4   & 7.4   & 11.4 \\
                     & 10    & 3.3   & 8.9   & 15.3  & 2.4   & 6.6   & 12.3  & 1.6   & 5.8   & 10.9 \\[8pt]
      \midrule               
      Sample Size & $r$  & \multicolumn{3}{c}{200} & \multicolumn{3}{c}{300} & \multicolumn{3}{c}{400} \\
      \cmidrule(lr){3-5}\cmidrule(lr){6-8}\cmidrule(lr){9-11}
      Nominal Size & & 1.0 & 5.0 & 10.0  & 1.0 & 5.0 & 10.0  & 1.0 & 5.0 & 10.0  \\
      \midrule
     $\xi=-0.25$ 	 & 1     & 5.4   & 10.5  & 15.2  & 3.6   & 8.2   & 12.6  & 2.8   & 7.1   & 12.5 \\
                     & 2     & 1.4   & 6.7   & 12.8  & 1.4   & 6.4   & 11.4  & 1.4   & 5.1   & 10.9 \\
                     & 3     & 1.5   & 5.9   & 11.8  & 1.1   & 5.4   & 10.8  & 1.2   & 6.6   & 11.9 \\
                     & 4     & 1.1   & 5.6   & 10.6  & 1.0   & 5.6   & 11.5  & 1.0   & 4.7   & 9.0 \\
                     & 5     & 1.1   & 4.5   & 9.3   & 1.2   & 5.7   & 11.7  & 1.2   & 4.7   & 10.2 \\
                     & 10    & 1.4   & 6.4   & 11.6  & 1.7   & 6.2   & 11.3  & 0.8   & 5.0   & 10.1 \\
    $\xi=0$  		 & 1     & 1.3   & 6.1   & 11.2  & 0.8   & 5.2   & 10.0  & 1.0   & 5.1   & 11.4 \\
                     & 2     & 0.5   & 5.0   & 10.6  & 1.2   & 5.7   & 11.8  & 1.0   & 5.9   & 11.0 \\
                     & 3     & 1.4   & 4.5   & 9.8   & 1.3   & 6.0   & 9.6   & 0.9   & 4.4   & 8.3 \\
                     & 4     & 1.1   & 5.4   & 10.6  & 1.3   & 5.2   & 9.9   & 0.9   & 5.0   & 9.1 \\
                     & 5     & 1.8   & 6.2   & 12.5  & 0.9   & 4.6   & 9.8   & 1.2   & 4.6   & 9.0 \\
                     & 10    & 1.1   & 3.8   & 9.3   & 0.9   & 5.2   & 12.6  & 1.2   & 4.9   & 9.7 \\
    $\xi=0.25$ 		 & 1     & 0.9   & 4.9   & 11.4  & 0.9   & 5.0   & 10.8  & 0.7   & 5.2   & 9.2 \\
                     & 2     & 3.2   & 7.9   & 11.7  & 2.3   & 7.1   & 11.2  & 2.5   & 6.6   & 12.1 \\
                     & 3     & 1.8   & 6.1   & 10.7  & 2.6   & 7.0   & 11.2  & 1.0   & 4.8   & 10.6 \\
                     & 4     & 1.2   & 4.9   & 11.2  & 1.2   & 6.0   & 9.9   & 1.2   & 5.8   & 11.8 \\
                     & 5     & 1.6   & 5.9   & 10.0  & 1.3   & 7.3   & 11.8  & 1.2   & 3.9   & 8.4 \\
                     & 10    & 1.7   & 6.6   & 12.4  & 0.9   & 4.4   & 9.8   & 1.6   & 5.7   & 10.4 \\
    \bottomrule
  \end{tabular}
  \label{tab:multsize}
\end{table}

To assess the convergence of $T_n^{(r)}(\hat\theta_n)$ to $N(0, 1)$,
10,000 replicates of the GEV$_r$ distribution were simulated with 
$\mu=0$ and $\sigma=1$ for each configuration of $r \in \{2,5,10\}$, 
$\xi \in \{-0.25, 0, 0.25\}$, and $n \in \{50, 100\}$. A rejection for 
nominal level $\alpha$, is denoted if 
$|T_n^{(r)}(\hat\theta_n)| > |Z_{\frac{\alpha}{2}}|$, 
where $Z_{\frac{\alpha}{2}}$ is the $\alpha/2$ percentile of the N(0,1) 
distribution. Using this result, the empirical size of the ED test can 
be summarized, and the results are presented in Table~\ref{tab:edsize}.

\begin{table}[tbp]
  \centering
  \caption{Empirical size (in \%) for the entropy difference (ED) 
  	test under the null distribution GEV$_r$, with $\mu=0$ and
    $\sigma=1$ based on 10,000 samples.}
    \begin{tabular}{cc rrr rrr}
      \toprule
      Sample Size & $r$ & \multicolumn{3}{c}{50} & \multicolumn{3}{c}{100} \\
      \cmidrule(lr){3-5}\cmidrule(lr){6-8}
      Nominal Size & & 1.0 & 5.0 & 10.0  & 1.0 & 5.0 & 10.0  \\
      \midrule
   $\xi = -0.25$ & 2     & 1.5   & 5.7   & 10.8  & 1.3   & 5.5   & 10.1 \\
          		 & 5     & 2.4   & 6.8   & 11.9  & 1.6   & 5.9   & 10.6 \\
         		 & 10    & 2.3   & 6.8   & 11.7  & 1.9   & 6.0   & 11.1 \\[6pt]
   $\xi = 0$     & 2     & 1.3   & 5.6   & 11.0  & 1.2   & 5.3   & 10.4 \\
          		 & 5     & 1.6   & 5.9   & 11.2  & 1.5   & 5.7   & 10.6 \\
          		 & 10    & 2.3   & 6.5   & 11.8  & 1.6   & 5.9   & 10.7 \\[6pt]
   $\xi = 0.25$  & 2     & 1.3   & 5.7   & 10.7  & 1.3   & 5.4   & 10.5 \\
          		 & 5     & 1.6   & 5.8   & 11.5  & 1.3   & 5.6   & 10.2 \\
          		 & 10    & 2.0   & 6.6   & 11.9  & 1.4   & 5.5   & 10.4 \\
    \bottomrule
    \end{tabular}
  \label{tab:edsize}
\end{table}

For sample size 50, the empirical size is above the nominal level 
for all configurations of $r$ and $\xi$. As the sample size increases 
from 50 to 100, the empirical size stays the same or decreases in 
every setting. For sample size 100, the agreement between nominal 
and observed size appears to be satisfactory for all configurations 
of $r$ and $\xi$. For sample size 50, the empirical size is 
slightly higher than the nominal size, but may be acceptable to some 
practitioners. For example, the empirical size for nominal size 10\% 
is never above 12\%, and for nominal size 5\%, empirical size is never 
above 7\%.

In summary, the multiplier bootstrap procedure of the 
score test can be used as a fast, reliable alternative to the parametric 
bootstrap procedure for sample size 100 or more when $\xi \ge 0$. 
When only small samples are available (less than 50 observations), the 
parametric bootstrap procedure is most appropriate since the multiplier 
version does not hold its size and the ED test relies upon samples of
size 50 or more for the central limit theorem to take effect.

\subsection{Power}
\label{ss:power}

The powers of the score tests and the ED test are studied with
two data generating schemes under the alternative hypothesis.
In the first scheme, 4~largest order statistics were generated from 
the GEV$_4$ distribution with $\mu = 0$, $\sigma = 1$, and 
$\xi \in \{-0.25, 0, 0.25\}$, and the 5th one was generated from a KumGEV 
distribution right truncated by the 4th largest order statistic.
The KumGEV distribution is a generalization of the GEV distribution
\citep{eljabri2013new} with two additional parameters $a$ and $b$ 
which alter skewness and kurtosis.
Defining $G_r({\bf x})$ to be the distribution function of the 
GEV$_r$($\mu, \sigma, \xi$) distribution, the distribution function
of the KumGEV$_r$($\mu, \sigma, \xi, a, b$) is given by 
$F_r({\bf x}) = 1 - \{1 - [G_r({\bf x})]^a\}^b$ for $a>0$, $b>0$.
The score test and the ED test were applied to the top~5 
order statistics with sample size $n \in \{100, 200\}$.
When $a = b = 1$, the null hypothesis of GEV$_5$ is true.
Larger difference from 1 of parameters $a$ and $b$ means 
larger deviation from the null hypothesis of GEV$_5$.

\begin{table}
  \centering
  \caption{Empirical rejection rate (in \%) of the multiplier score test and the ED test in the first data generating scheme from 1000 replicates.}
  \begin{tabular}{c rr rrrrrrrrr}
    \toprule
    Sample Size & $\xi$   & Test & \multicolumn{9}{c}{Value of $a$=$b$} \\
    \cmidrule(lr){4-12}
    &  &   & 0.4   & 0.6   & 0.8   & 1.0     & 1.2   & 1.4   & 1.6   & 1.8   & 2.0 \\
    \midrule
    100   & $-$0.25 & Score & 99.9  & 84.8  & 20.4  & 5.4   & 21.0  & 41.2  & 62.3  & 79.0  & 83.0 \\
          &         & ED   & 100.0 & 99.0  & 46.5  & 4.6   & 48.7  & 89.5  & 99.2  & 100.0 & 99.8 \\
          & 0       & Score & 100.0 & 87.0  & 21.6  & 7.4   & 24.2  & 48.9  & 67.8  & 79.6  & 89.4 \\
          &         & ED   & 100.0 & 98.8  & 40.0  & 5.2   & 40.6  & 87.2  & 98.5  & 100.0 & 99.7 \\
          & 0.25    & Score & 100.0 & 87.7  & 20.3  & 6.2   & 25.8  & 54.2  & 74.2  & 82.9  & 89.5 \\
          &         & ED   & 100.0 & 97.5  & 37.7  & 4.8   & 34.8  & 78.1  & 96.1  & 99.5  & 99.7 \\[6pt]
    200   & $-$0.25 & Score & 100.0 & 98.6  & 40.7  & 5.2   & 29.8  & 64.7  & 86.4  & 95.9  & 97.5 \\
          &         & ED   & 100.0 & 100.0 & 78.4  & 6.2   & 70.0  & 99.2  & 100.0 & 100.0 & 100.0 \\
          & 0       & Score & 100.0 & 99.4  & 44.6  & 6.1   & 34.9  & 75.0  & 92.4  & 97.3  & 98.6 \\
          &         & ED   & 100.0 & 99.9  & 75.0  & 5.5   & 64.6  & 98.1  & 99.8  & 100.0 & 100.0 \\
          & 0.25    & Score & 100.0 & 99.3  & 44.5  & 6.3   & 37.0  & 73.4  & 91.8  & 97.0  & 98.9 \\
          &         & ED   & 100.0 & 100.0 & 71.0  & 5.2   & 57.2  & 95.9  & 100.0 & 100.0 & 100.0 \\
          \bottomrule
    \end{tabular}
  \label{tab:kum}
\end{table}

Table~\ref{tab:kum} summarizes the empirical rejection 
percentages obtained with nominal size 5\%, 
for a sequence value of $a = b$ from 0.4 to 2.0, with increment 0.2.
Both tests hold their sizes when $a = b = 1$ and have substantial 
power in rejecting the null hypothesis for other values of $a = b$.
Between the two tests, the ED test demonstrated much higher power 
than the score test in the more difficult cases where the deviation
from the null hypothesis is small; for example, the ED test's power
almost doubled the score test's power for $a = b \in \{0.8, 1.2\}$.
As expected, the powers of both tests increase as $a = b$ moves
away from 1 or as the sample sizes increases.

In the second scheme, top~6 order statistics were generated
from the GEV$_{6}$ distribution with $\mu=0$, $\sigma=1$, and 
$\xi \in \{ -0.25, 0, 0.25\}$, and then the 5th order statistic
was replaced from a mixture of the 5th and 6th order statistics.
The tests were applied to the sample of first~5 order 
statistics with sample sizes $n \in \{100, 200\}$.
The mixing rate $p$ of the 5th order statistic took
values in $\{0.00, 0.10, 0.25, 0.50, 0.75, 0.90, 1.00\}$.
When $p = 1$ the null hypothesis of GEV$_5$ is true.
Smaller values of $p$ indicate larger deviations from the null.
Again, both tests hold their sizes when $p = 1$ and have 
substantial power for other values of $p$, which increases 
as $p$ decreases or as the sample sizes increases.
The ED test again outperforms the score test with almost
doubled power in the most difficult cases with $p \in \{0.75, 0.90\}$.
For sample size 100 with $p = 0.50$, for instance, the ED test 
has power above 93\% while the score test only has power above 69\%.

\begin{table}
  \centering
  \caption{Empirical rejection rate (in \%) of the multiplier score test and the ED tests in the second data generating scheme from 1000 replicates.}
  \begin{tabular}{c rr rrrrrrr}
    \toprule
    Sample Size &  $\xi$   & Test & \multicolumn{7}{c}{Mixing Rate $p$} \\
    \cmidrule(lr){4-10}
    & & &    0.00   & 0.10   & 0.25   & 0.50   & 0.75   & 0.90   & 1.00 \\
    \midrule
    100   & $-$0.25 & Score & 99.7  & 99.5  & 95.5  & 69.4  & 24.1  & 7.8   & 5.8 \\
          &         & ED    & 100.0 & 100.0 & 100.0 & 97.7  & 51.8  & 10.9  & 6.2 \\
          & 0       & Score & 100.0 & 99.7  & 97.8  & 72.4  & 22.7  & 6.2   & 6.8 \\
          &         & ED    & 100.0 & 100.0 & 100.0 & 96.0  & 47.6  & 10.3  & 5.6 \\
          & 0.25    & Score & 99.9  & 99.7  & 96.6  & 70.8  & 24.7  & 5.8   & 5.3 \\
          &         & ED    & 100.0 & 100.0 & 99.9  & 93.6  & 43.4  & 9.8   & 5.2 \\[6pt]
    200   & $-$0.25 & Score & 99.9  & 100.0 & 99.7  & 95.6  & 43.4  & 11.4  & 5.1 \\
          &         & ED    & 100.0 & 100.0 & 100.0 & 100.0 & 83.6  & 20.0  & 5.8 \\
          & 0       & Score & 100.0 & 100.0 & 100.0 & 96.5  & 44.4  & 11.2  & 5.4 \\
          &         & ED    & 100.0 & 100.0 & 100.0 & 100.0 & 79.5  & 20.0  & 5.5 \\
          & 0.25    & Score & 100.0 & 100.0 & 100.0 & 97.2  & 46.9  & 9.2   & 5.5 \\
          &         & ED    & 100.0 & 100.0 & 100.0 & 99.7  & 72.5  & 17.9  & 4.2 \\
          \bottomrule
    \end{tabular}
  \label{tab:mix1}
\end{table}

\section{Automated Sequential Testing Procedure}
\label{s:seq}

As there are $R$ hypotheses $H_0^{(r)}$, $r = 1, \ldots, R$,
to be tested in a sequence in the methods proposed, the 
sequential, multiple testing issue needs to be addressed.
Most methods for error control assume that all the tests can 
be run first and then a subset of tests are chosen to be rejected
\citep[e.g.,][]{benjamini2010discovering,benjamini2010simultaneous}.
The errors to be controlled are either the 
FWER \citep{shaffer1995multiple}, or the 
FDR \citep{Benjamini1995,BY2001}.
In contrast to the usual multiple testing procedures, however,
a unique feature in this setting is that the hypotheses must
be rejected in an ordered fashion: if $H_0^{(r)}$ is rejected,
$r < R$, then $H_0^{(k)}$ will be rejected for all $r < k \le R$.
Despite the extensive literature on multiple testing and
the more recent developments on FDR control and its variants, 
no definitive procedure has been available for error control in 
ordered tests until the recent work of \citet{g2015sequential}.

Consider a sequence of null hypotheses $H_1, \ldots, H_m$.
An ordered test procedure must reject $H_1, \ldots, H_k$
for some $k \in \{0, 1, \ldots, m\}$, which rules out the
classical methods for FDR control \citep{Benjamini1995}.
Let $p_1, \ldots, p_m \in [0, 1]$ be the corresponding 
p-values of the $m$ hypotheses such that $p_j$ is 
uniformly distributed over $[0,1]$ when $H_j$ is true. 
The methods of \citet{g2015sequential} transform the sequence
of p-values to a monotone sequence and then apply the original
Benjamini--Hochberg procedure on the monotone sequence.
They proposed two rejections rules, each returning a cutoff 
$\hat k$ such that $H_1, \ldots, H_{\hat k}$ are rejected.
The first is called ForwardStop,
\begin{equation*}
\hat{k}_{\mathrm{F}} = \max \left\{k \in \{1, \ldots, m\}: -\frac{1}{k} \sum_{i=1}^k \log(1-p_i) \leq \alpha \right\},
\end{equation*}
and the second is called StrongStop,
\begin{equation*}
\hat{k}_{\mathrm{S}} = \max \left\{k \in \{1, \ldots, m\}:
  \exp\Big(\sum_{j=k}^m \frac{\log p_j}{j}\Big)   \leq \frac{\alpha
    k}{m} \right\},
\end{equation*}
where $\alpha$ is a pre-specified level. Both rules were 
shown to control the FDR at level $\alpha$ 
under the assumption of independent p-values.
ForwardStop sets the rejection threshold at the largest $k$ at which
the average of first $k$ transformed p-values is small enough.
As it does not depend on those p-values with later indices, 
this rule is robust to potential misspecification at later indices.
StrongStop offers a stronger guarantee than ForwardStop. 
If the non-null p-values indeed precede the null p-values, 
it controls the FWER at level $\alpha$ in addition to the FDR. 
Thus, for ForwardStop, this $\alpha$ refers to the FDR and for 
StrongStop, $\alpha$ refers to the FWER. 
As the decision to stop at $k$ depends on all the p-values after $k$,
its power may be harmed if, for example, the very last p-values
are slightly higher than expected under the null hypotheses.

To apply the two rules to our setting, note that our objective 
is to give a threshold $\hat r$ such that the first $\hat r$ 
of $m = R$ hypotheses are accepted instead of rejected.
Therefore, we put the p-values in reverse order:
let the ordered set of p-values $\{p_1, \ldots, p_R\}$ 
correspond to hypotheses $\{H_0^{(R)}, \ldots, H_0^{(1)}\}$.
The two rules give a cutoff $\hat{k} \in \{1, \ldots, R\}$ 
such that the hypotheses 
$H_0^{(R)}, \ldots, H_0^{(R - \hat{k} + 1)}$ are rejected.
If no $\hat{k} \in \{1, \ldots, R\}$ exists, then no rejection is made.

A caveat is that, unlike the setting of~\cite{g2015sequential}, 
the p-values of the sequential tests are dependent. 
Nonetheless, the ForwardStop and StrongStop procedures 
may still provide some error control. For example, in the 
non-sequential multiple testing scenario~\cite{BY2001} 
show that their procedure controls the FDR under certain 
positive dependency conditions, while~\cite{blanchard2009adaptive} 
implement adaptive versions of step-up procedures that 
provably control the FDR under unspecified dependence 
among p-values.

\begin{figure*}[tbp]
    \centering
      \includegraphics[width=\textwidth]{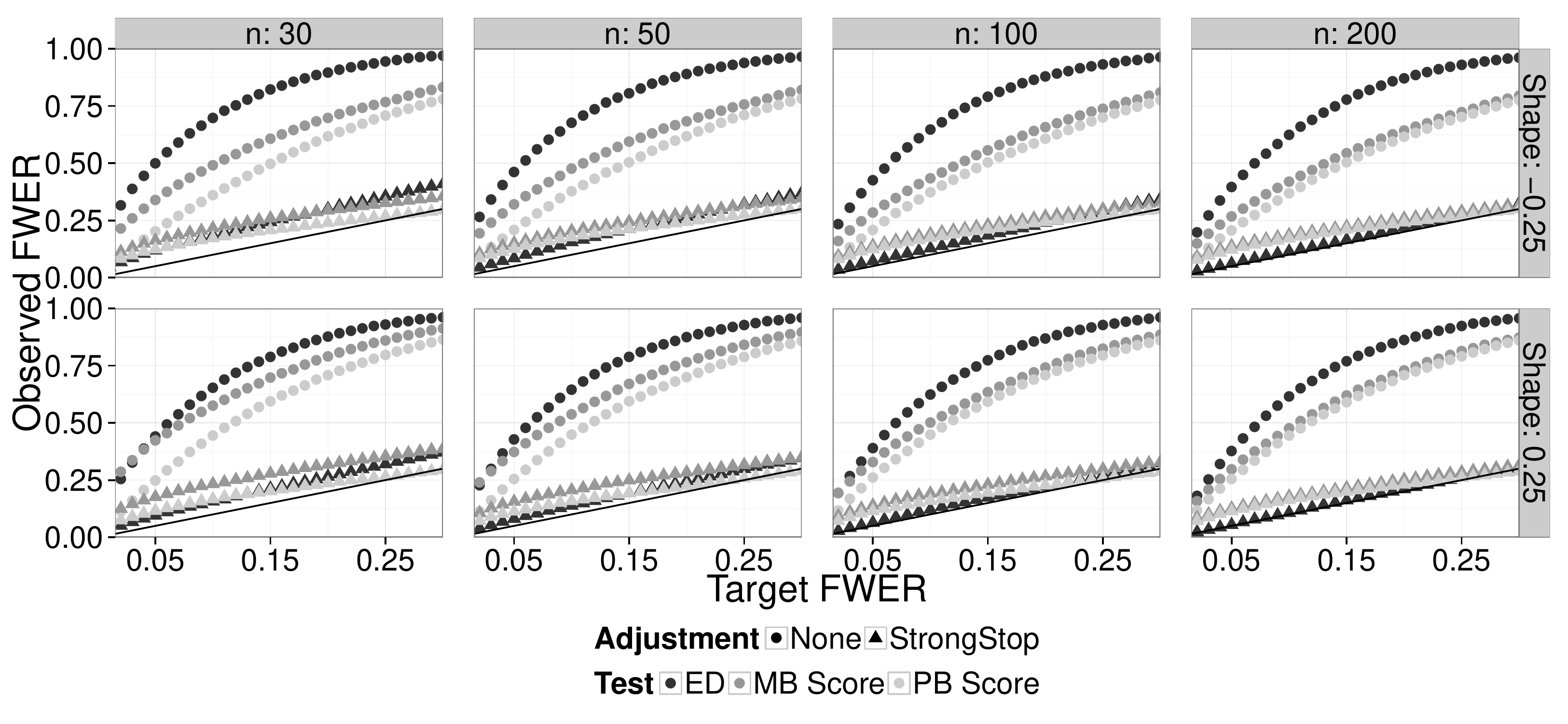}
    \caption{
      Observed FWER for the ED, parametric bootstrap (PB) score, and
      multiplier bootstrap (MB) score tests (using No Adjustment and StrongStop) 
      versus expected FWER at various nominal levels. The 45 degree line indicates 
      agreement between the observed and expected rates under $H_0$.}
    \label{fig:FWERcontrol}
\end{figure*}

The empirical properties of the two rules for the tests 
in this paper are investigated in simulation studies.
To check the empirical FWER of the StrongStop rule, 
only data under the null hypotheses are needed.
With $R=10$, $\xi \in \{-0.25, 0.25\}$, $n \in \{30, 50, 100, 200\}$, 
$\mu=0$, and $\sigma=1$, 1000 GEV$_{10}$ samples were generated. 
For the ED, multiplier bootstrap score, and parametric bootstrap score
test, the observed FWER is compared to the expected rates at various 
nominal $\alpha$ control levels. The StrongStop procedure is 
used, as well as no error control (i.e. a rejection occurs any 
time the raw p-value is below the nominal level). The results of 
this simulation are presented in Figure~\ref{fig:FWERcontrol}.

It is clear that the StrongStop reasonably controls the FWER 
for the ED test and the agreement between the observed 
and expected rate increases as the sample size increases. 
For both the parametric and multiplier bootstrap 
versions of the score test however, the observed FWER is above 
the expected rate, at times 10\% higher. Regardless, it is 
apparent that using no error control results in an inflated 
FWER, and this inflation can only increase as the number of 
tests increase.

To check the empirical FDR of the ForwardStop rule, 
data need to be generated from a non-null model.
To achieve this, consider the sequence of specification tests 
of GEV$_r$ distribution with $r \in \{1, \ldots, 6\}$, 
where the 5th and 6th order statistics are misspecified. 
Specifically, data from the GEV$_7$ distribution with 
$\mu=0$ and $\sigma=1$ were generated for $n$ blocks;
then the 5th order statistic is replaced with a 50/50 mixture of 
the 5th and 6th order statistics, and the 6th order statistic is 
replaced with a 50/50 mixture of the 6th and 7th order statistics. 
This is replicated 1000 times for each value of 
$\xi \in \{-0.25, 0.25\}$ and $n \in \{30, 50, 100, 200\}$. 
For nominal level $\alpha$, the observed FDR is defined as the number
of false rejections (i.e. any rejection of $r \leq 4$) divided by the
number of total rejections.

\begin{figure*}[tbp]
\centering
\includegraphics[width=\textwidth]{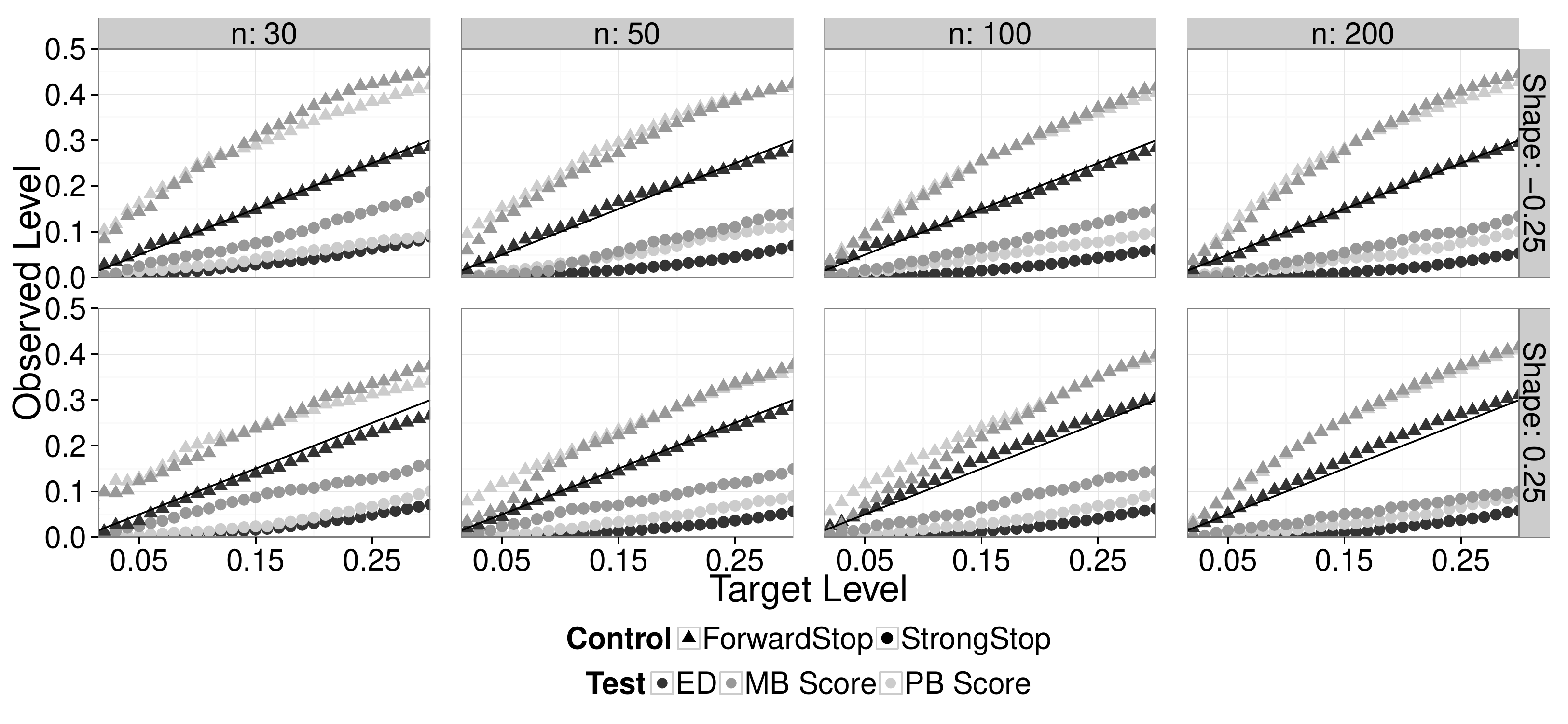}
\caption{Observed FDR (from ForwardStop) and observed FWER (from StrongStop) 
  versus expected FDR and FWER, respectively, at various nominal levels. 
  This is for the simulation setting described in Section~\ref{s:seq}, 
  using the ED, parametric bootstrap (PB) score, and multiplier bootstrap
  (MB) score tests. The 45 degree line indicates agreement between the 
  observed and expected rates.}
\label{fig:ChoiceOfR}
\end{figure*}

The results are presented in Figure~\ref{fig:ChoiceOfR}. 
The plots show that the ForwardStop 
procedure controls the FDR for the ED test, while for both 
versions of the score test, the observed FDR is slightly 
higher than the expected at most nominal rates. Here, sample 
size does not appear to effect the observed rates.

Similarly, the observed FWER rate in this particular simulation setting 
can be found by taking the number of simulations with at least one 
false rejection (here, any rejection of $r \leq 4$) and dividing that number 
by the total number of simulations. This calculation is performed for a 
variety of nominal levels $\alpha$, using the StrongStop procedure. 
The results are presented in Figure~\ref{fig:ChoiceOfR}. In 
this particular simulation setting, the StrongStop procedure controls the 
FWER for the ED test and both versions of the score test at all sample 
sizes investigated.

It is of interest to investigate the performance of the ForwardStop
and StrongStop in selecting $r$ for the $r$ largest order statistics method.
In the simulation setting described in the last paragraph, the
correct choice of $r$ should be 4, and a good testing procedure 
should provide a selection $\hat r$ close to 4.
The choice $\hat r \in \{0, \ldots, 6\}$ is recorded using the ED test
and bootstrap score tests with both ForwardStop and StrongStop. 
Due to space constraints, we choose to present one setting, 
where $\xi=0.25$ and $n=100$. The non-adjusted sequential procedure 
is also included, testing in an ascending manner from $r = 1$ and 
$\hat r$ is chosen by the first rejection found (if any). 
The results are summarized in Table~\ref{tab:multcomp}.

\begin{table}[tbp]
  \centering
  \footnotesize
  \caption{Percentage of choice of $r$ using the ForwardStop and
    StrongStop rules at various significance levels or FDRs, under ED,
    parametric bootstrap (PB) score, and multiplier bootstrap (MB)
    score tests,  with $n=100$ and $\xi=0.25$ for the simulation setting
    described in Section~\ref{s:seq}. Correct choice is $r=4$. }
  \rotatebox{90}{
    \begin{tabular}{rc rrrrrr rrrrrr rrrrrr}
    \toprule
    Test  & r     & \multicolumn{6}{c}{Unadjusted}                       & \multicolumn{6}{c}{ForwardStop}                      & \multicolumn{6}{c}{StrongStop} \\
    \cmidrule(lr){3-8} \cmidrule(lr){9-14}  \cmidrule(lr){15-20} 
      & Significance:  & 0.01  & 0.05  & 0.1   & 0.2  & 0.3 & 0.4 & 0.01  & 0.05  & 0.1   & 0.2  & 0.3 & 0.4 & 0.01  & 0.05  & 0.1   & 0.2  & 0.3 & 0.4 \\
         \midrule
    ED    & 6       & 19.0  & 3.4   & 1.5   & 0.5   & 0.0   & 0.0   & 86.6  & 69.8  & 58.5  & 43.0  & 30.0  & 22.4  & 52.4  & 22.2  & 13.1  & 5.4   & 1.7   & 1.0 \\
          & 5       & 1.9   & 2.1   & 1.1   & 0.7   & 0.2   & 0.1   & 1.3   & 1.5   & 1.1   & 0.9   & 0.2   & 0.0   & 25.0  & 18.9  & 13.6  & 6.7   & 2.9   & 1.4 \\
       \rowcolor{lightgray} & 4     & 76.2  & 79.9  & 70.2  & 50.5  & 35.2  & 22.3  & 12.0  & 25.1  & 31.7  & 33.4  & 31.6  & 26.3  & 22.6  & 58.9  & 72.9  & 84.9  & 89.0  & 85.7 \\
          & 3       & 0.8   & 4.1   & 7.6   & 11.8  & 15.3  & 15.1  & 0.1   & 3.4   & 6.2   & 12.4  & 17.0  & 18.5  & 0.0   & 0.0   & 0.3   & 2.2   & 4.5   & 6.9 \\
          & 2       & 1.1   & 5.3   & 9.5   & 16.2  & 19.9  & 22.8  & 0.0   & 0.1   & 1.9   & 5.4   & 9.4   & 11.5  & 0.0   & 0.0   & 0.1   & 0.7   & 1.8   & 4.4 \\
          & 1       & 1.0   & 5.2   & 10.1  & 20.3  & 29.4  & 39.7  & 0.0   & 0.1   & 0.6   & 4.9   & 11.8  & 21.3  & 0.0   & 0.0   & 0.0   & 0.1   & 0.1   & 0.6 \\
          & 0       & -     & -     & -     & -     & -     & -     & -     & -     & -     & -     & -     & -     & -     & -     & -     & -     & -     & - \\[6pt]
  PB Score & 6       & 35.8  & 16.1  & 8.5   & 1.8   & 0.6   & 0.2   & 53.5  & 33.0  & 23.4  & 13.9  & 8.6   & 5.9   & 40.6  & 25.1  & 18.8  & 12.1  & 7.6   & 5.6 \\
          & 5       & 2.5   & 1.8   & 0.8   & 0.5   & 0.2   & 0.1   & 1.9   & 1.3   & 0.7   & 0.8   & 0.4   & 0.4   & 29.8  & 37.7  & 29.2  & 17.5  & 10.5  & 6.6 \\
       \rowcolor{lightgray} & 4     & 58.4  & 68.1  & 63.9  & 46.1  & 31.3  & 19.8  & 42.6  & 57.8  & 58.9  & 51.1  & 42.0  & 31.1  & 29.3  & 36.5  & 50.5  & 65.7  & 73.5  & 74.6 \\
          & 3      & 0.5    & 2.3   & 4.5   & 6.8   & 7.1   & 8.0   & 1.4   & 4.2   & 9.0   & 15.5  & 16.0  & 17.1  & 0.0   & 0.4   & 1.1   & 3.3   & 4.8   & 5.6 \\
          & 2      & 0.6    & 3.0   & 4.9   & 10.3  & 11.7  & 12.7  & 0.5   & 1.7   & 3.0   & 7.3   & 10.2  & 12.3  & 0.1   & 0.1   & 0.2   & 1.0   & 2.4   & 4.5 \\
          & 1      & 0.8    & 3.7   & 6.9   & 13.9  & 18.3  & 18.5  & 0.1   & 1.3   & 2.3   & 4.6   & 8.3   & 9.6   & 0.2   & 0.2   & 0.2   & 0.4   & 1.2   & 2.7 \\
          & 0      & 1.4    & 5.0   & 10.5  & 20.6  & 30.8  & 40.7  & 0.0   & 0.7   & 2.7   & 6.8   & 14.5  & 23.6  & 0.0   & 0.0   & 0.0   & 0.0   & 0.0   & 0.4 \\[6pt]
  MB Score & 6      & 49.9   & 16.9  & 6.9   & 1.3   & 0.2   & 0.0   & 71.7  & 40.3  & 24.7  & 12.6  & 7.8   & 5.5   & 51.6  & 27.3  & 16.9  & 10.4  & 6.2   & 4.3 \\
          & 5      & 2.5    & 2.3   & 0.7   & 0.3   & 0.1   & 0.0   & 1.3   & 2.0   & 0.8   & 0.5   & 0.3   & 0.4   & 39.4  & 50.7  & 42.9  & 25.5  & 15.8  & 9.4 \\
      \rowcolor{lightgray}  & 4     & 38.3  & 59.6  & 59.1  & 44.0  & 31.2  & 18.4  & 26.6  & 53.3  & 59.3  & 49.9  & 40.1  & 28.0  & 6.2   & 18.5  & 35.0  & 55.4  & 62.3  & 64.8 \\
          & 3      & 1.6    & 2.8   & 4.0   & 6.6   & 7.5   & 6.0   & 0.3   & 2.8   & 7.4   & 15.7  & 16.0  & 17.9  & 0.6   & 1.2   & 2.5   & 3.6   & 7.0   & 7.8 \\
          & 2      & 2.7    & 4.4   & 7.1   & 11.0  & 10.0  & 10.6  & 0.1   & 0.6   & 3.4   & 8.6   & 9.7   & 9.5   & 0.7   & 0.8   & 1.2   & 2.8   & 5.3   & 7.0 \\
          & 1      & 4.2    & 8.3   & 10.6  & 14.0  & 19.5  & 20.0  & 0.0   & 0.9   & 2.7   & 4.7   & 7.9   & 7.8   & 1.5   & 1.5   & 1.5   & 2.3   & 3.4   & 6.4 \\
          & 0      & 0.8    & 5.7   & 11.6  & 22.8  & 31.5  & 45.0  & 0.0   & 0.1   & 1.7   & 8.0   & 18.2  & 30.9  & 0.0   & 0.0   & 0.0   & 0.0   & 0.0   & 0.3 \\
    \bottomrule
    \end{tabular}
  \label{tab:multcomp}
  }
\end{table}

In general, larger choices of $\alpha$ lead to a higher percentage of
$\hat r=4$ being correctly chosen with ForwardStop or StrongStop.
Intuitively, this is not surprising since a smaller $\alpha$ makes it 
more difficult to reject the `bad' hypotheses of $r\in \{5,6\}$. 
A larger choice of $\alpha$ also leads to a higher probability 
of rejecting too many tests; i.e. choosing $r$ too small. 
From the perspective of model specification, 
this is more desirable than accepting true negatives.  
A choice of 6, 5, or 0 is problematic, but choosing 1, 2, 
or 3 is acceptable, although some information is lost.
When no adjustment is used and an ascending sequential 
procedure is used, both tests have reasonable classification rates.
When $\alpha = 0.05$, the ED test achieves the correct choice of $r$
79.9\% of the time, with the parametric bootstrap and multiplier
bootstrap score tests achieving 68.1\% 
and 59.6\% respectively. Of course, as the number of tests (i.e., $R$)
increase, with no adjustment the correct classification rates will 
go down and the ForwardStop/StrongStop procedures will achieve better rates. 
This may not be too big an issue here as $R$ is typically small.
In the case where rich data are available and $R$ is big, the
ForwardStop and StrongStop becomes more useful as they are
designed to handle a large number of ordered hypothesis.

\section{Illustrations}
\label{s:app}

\subsection{Lowestoft Sea Levels}
\label{ss:lowe}

Sea level readings in 60 and 15 minute intervals from a gauge 
at Lowestoft off the east coast of Britain during the years 
1964--2014 are available from the UK Tide Gauge Network website. 
The readings are hourly from 1964--1992 and in fifteen minute 
intervals from 1993 to present. Accurate estimates of extreme 
sea levels are of great interest. The current data are of better 
quality and with longer record than those used in 
\citet{tawn1988extreme} --- annual maxima during 1953--1983 and 
hourly data during 1970--78 and 1980--82.

Justification of the statistical model was considered in detail 
by~\cite{tawn1988extreme}. The three main assumptions needed to
justify use of the GEV$_r$ model are: 
(1) The block size $B$ is large compared to the choice of $r$;
(2) Observations within each block and
across blocks are approximately independent; and 
(3) The distribution of the block maxima follows GEV$_1$.
The first assumption is satisfied, by letting $R = 125$, and noting that
the block size for each year is $B = 365 \times 24 = 8760$ from
1964--1992 and $B = 365 \times 96 = 35040$ from 1993--2014. 
This ensures that $r \ll B$. 
The third assumption is implicitly addressed in the testing procedure;
if the goodness-of-fit test for the block maxima rejects, all
subsequent tests for $r > 1$ are rejected as well.

The second assumption can be addressed in this setting by
the concept of independent storms~\citep{tawn1988extreme}. 
The idea is to consider each storm as a separate event, 
with each storm having some storm length, say $\tau$. 
Thus, when selecting the $r$ largest values from each block, 
only a single contribution can be obtained from each storm, which 
can be considered the $r$ largest independent annual events. 
By choosing $\tau$ large enough, this ensures both approximate
independence of observations within each block and across blocks. 
The procedure to extract the independent $r$ largest annual events is
as follows: 
\begin{enumerate}
\item
Pick out the largest remaining value from the year (block) of interest. 

\item
Remove observations within a lag of $\tau / 2$ from both sides of the value 
chosen in step~1.

\item
Repeat (within each year) until the $r$ largest are extracted. 
\end{enumerate}

\begin{figure*}[tbp]
    \includegraphics[width=\textwidth]{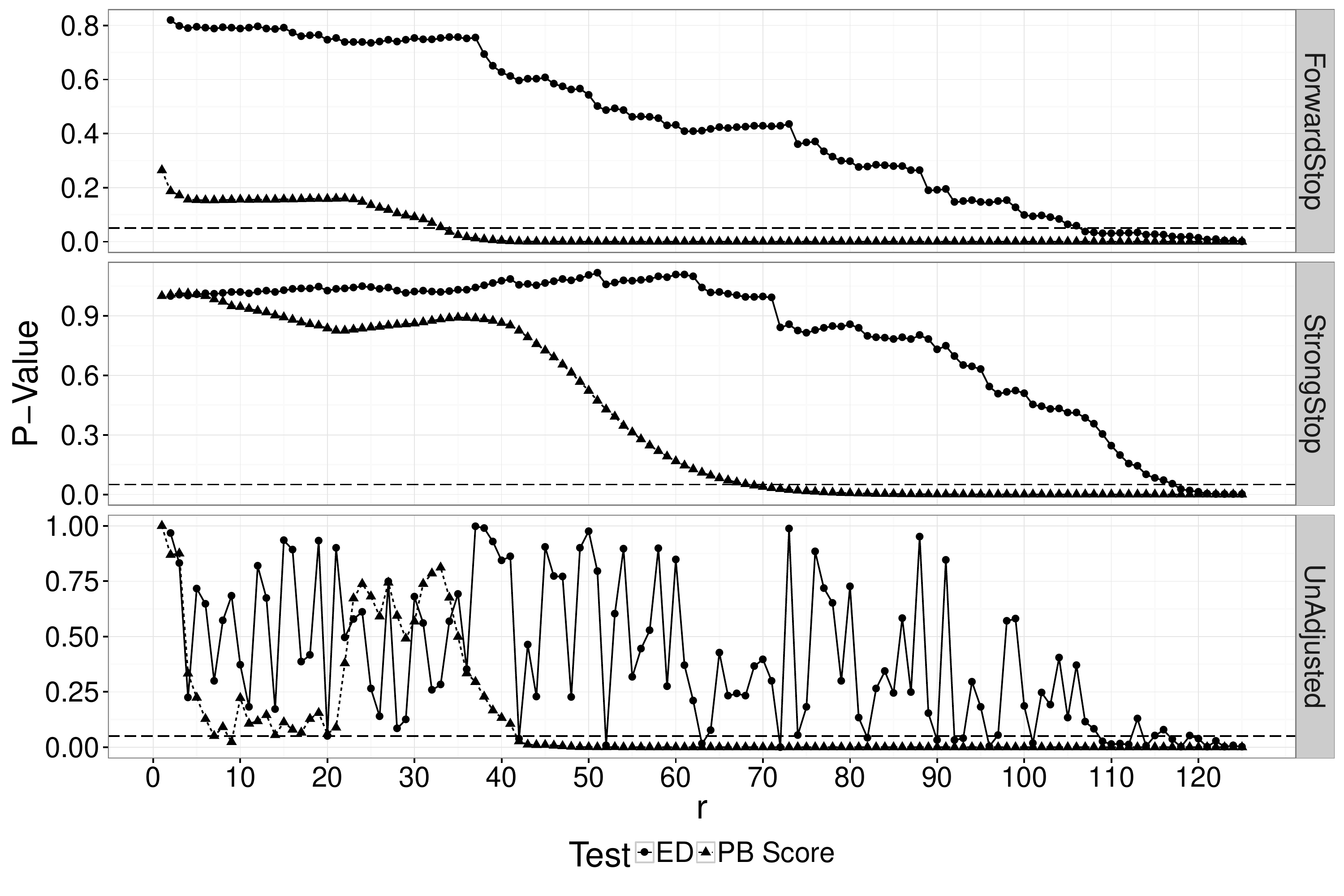}
    \caption{P-Values using ForwardStop, StrongStop, and no adjustment for 
    the ED and PB Score tests applied to the Lowestoft sea level data. 
    The horizontal dashed line represents the 0.05 possible cutoff value.}
    \label{fig:lowestoft_pvals}
\end{figure*}

\begin{figure*}[tbp]
  \includegraphics[width=\textwidth]{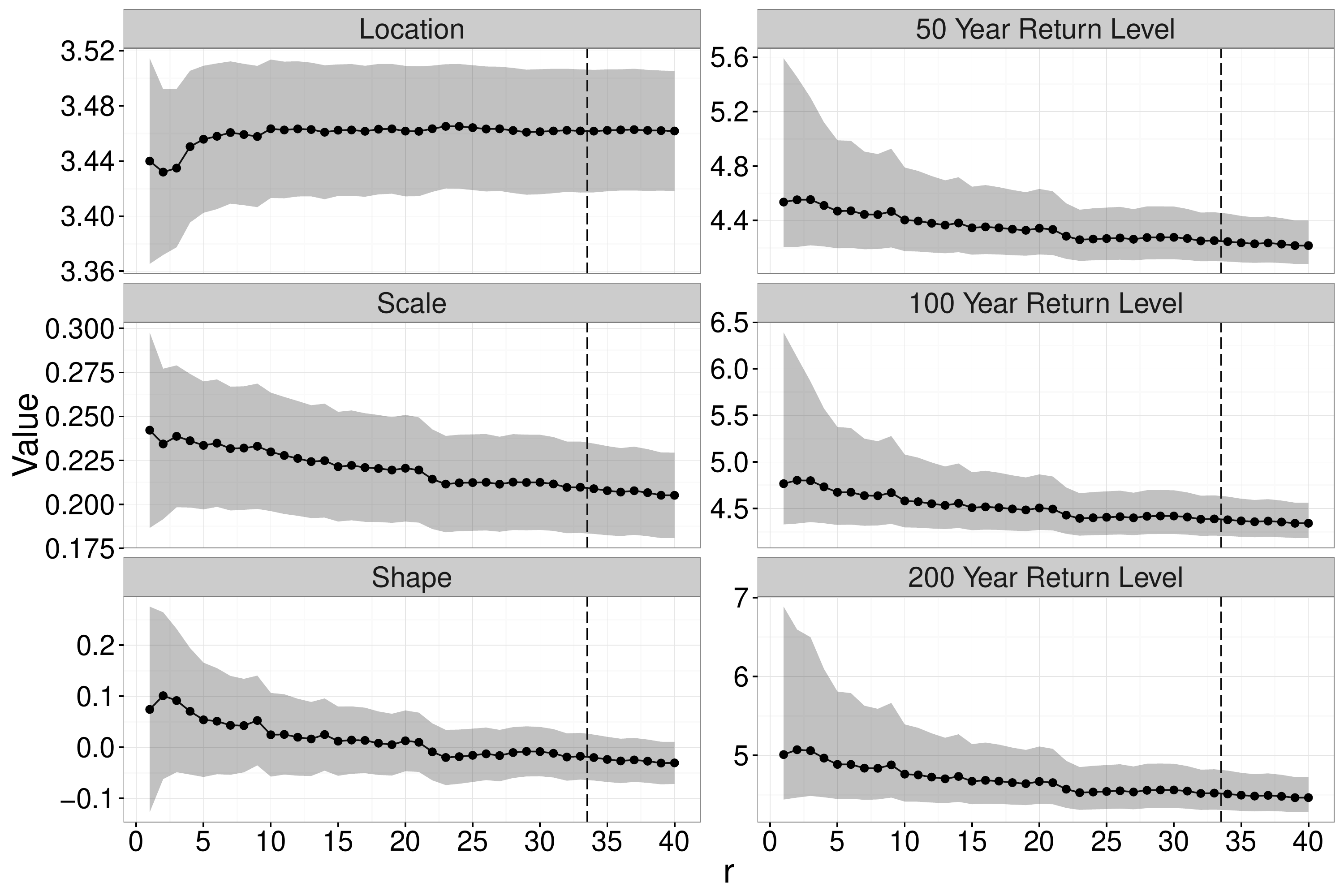}
  \caption{Location, scale, and shape parameter estimates, with 95\% profile
    confidence intervals for $r=1, \ldots, 40$ for the Lowestoft sea level data.
    Also included are the estimates and 95\% profile likelihood confidence
    intervals for the 50 and 100 year return levels.}
    \label{fig:lowestoft_params}
\end{figure*}

A full analysis is performed on the Lowestoft sea level data using 
$\tau = 60$ as the estimated storm length \citep{tawn1988extreme}. 
Using $R = 125$, both the parametric bootstrap score (with bootstrap
sample size $L = 10,000$) and ED test are applied sequentially on the
data. The p-values of the sequential 
tests (adjusted and unadjusted) can be seen in 
Figure~\ref{fig:lowestoft_pvals}. Due to the large number of tests, 
the adjustment for multiplicity is desired and thus, ForwardStop is 
used to choose $r$. For this dataset, the score test is more powerful 
than the ED test. With ForwardStop and the score test, 
Figure~\ref{fig:lowestoft_pvals} suggests that $r = 33$. 
The remainder of this analysis proceeds with the choice of $r = 33$. 
The estimated parameters and corresponding 95\% profile confidence intervals 
for $ r= 1$ through $r = 40$ are shown in Figure~\ref{fig:lowestoft_params}.

When $r = 33$, the parameters are estimated as $\hat{\mu} = 3.462\ (0.023)$, 
$\hat{\sigma}= 0.210\ (0.013)$, and $\hat{\xi}= -0.017\ (0.023)$, with
standard errors in parenthesis. 
An important risk measure is the $t$-year return level $z_t$
\citep[e.g.,][]{hosking1990moments,ribereau2008estimating,singo2012flood}.
It can be thought of here as the sea level that is exceeded once every 
$t$ years on average. Specifically, the $t$-year return level is 
the $1 - 1/t$ quantile of the GEV distribution
\[ z_t = 
\begin{cases} 
  \mu - \frac{\sigma}{\xi}\big\{1 - [ -  \log(1-\frac{1}{t})]^{-\xi} \big\}, & \xi \neq 0, \\
  \mu - \sigma\log[-\log(1-\frac{1}{t})], & \xi = 0.
\end{cases}
\]
The return levels can be estimated with parameter values 
replaced with their estimates, and confidence intervals 
can be constructed using profile likelihood
\citep[e.g.,][p.57]{coles2001introduction}.

The 95\% profile likelihood confidence intervals for the 50, 100,  
and 200 year return levels (i.e. $z_{50}, z_{100}, z_{200}$) are given by 
$(4.102, 4.461)$, $(4.210, 4.641)$ and $(4.312, 4.824)$, respectively. 
The benefit of using $r=1$ versus $r=33$ can be seen in the 
return level confidence intervals in Figure~\ref{fig:lowestoft_params}. 
For example, the point estimate of the 100 year return level decreases 
slightly as $r$ increases and the width of the 95\% confidence interval 
decreases drastically from 2.061 ($r=1$) to 0.432 ($r=33$), 
as more information is used. The lower bound of the interval 
however remains quite stable, shifting from 4.330 to 4.210 --- less 
than a 3\% change. Similarly, the standard error of the shape parameter 
estimate decreases by over two-thirds when using $r=33$ versus $r=1$. 

\subsection{Annual Maximum Precipitation: Atlantic City, NJ}
\label{ss:AC}

The top 10 annual precipitation events (in centimeters) were taken from 
the daily records of a rain gauge station in Atlantic City, NJ from 
1874--2015. The year 1989 is missing, while the remaining records 
are greater than 98\% complete. This provides a total record length 
of 141 years. The raw data is a part of the Global Historical Climatology 
Network (GHCN-Daily), with an overview given by~\cite{menne2012overview}. 
The specific station identification in the dataset is USW00013724.

\begin{figure*}[tbp]
    \includegraphics[width=\textwidth]{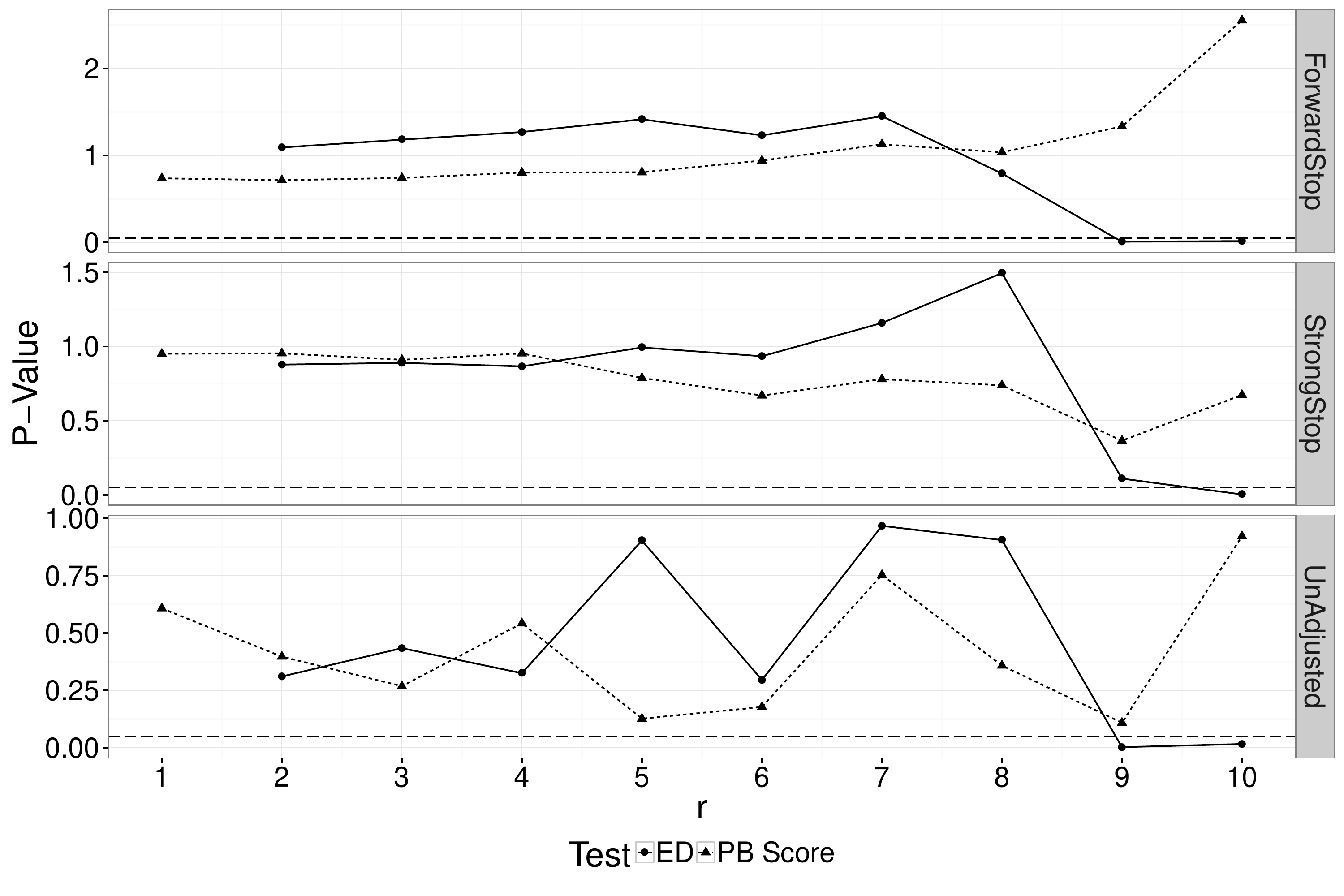}
    \caption{P-Values using ForwardStop, StrongStop, and no adjustment for 
    the ED and PB Score tests applied to the Atlantic City precipitation data. 
    The horizontal dashed line represents the 0.05 possible cutoff value.}
    \label{fig:AC_pvals}
\end{figure*}

Unlike for the Lowestoft sea level data, a rather small value is set for 
$R$ at $R = 10$ because of the much lower frequency of the daily data. 
Borrowing ideas from Section~\ref{ss:lowe}, a storm length of 
$\tau = 2$ is used to ensure approximate independence of observations.
Both the parametric bootstrap score (with $L = 10,000$) and ED test 
are applied sequentially on the data. 
The p-values of the sequential tests 
(ForwardStop, StrongStop, and unadjusted) are shown in 
Figure~\ref{fig:AC_pvals}. 
The score test does not pick up anything.
The ED test obtains p-values 0.002 and 0.016, respectively, 
for $r=9$ and $r=10$, which translates into a rejection 
using ForwardStop. Thus, Figure~\ref{fig:AC_pvals} suggests that $r=8$ 
be used for the analysis.

\begin{figure*}[tbp]
    \includegraphics[width=\textwidth]{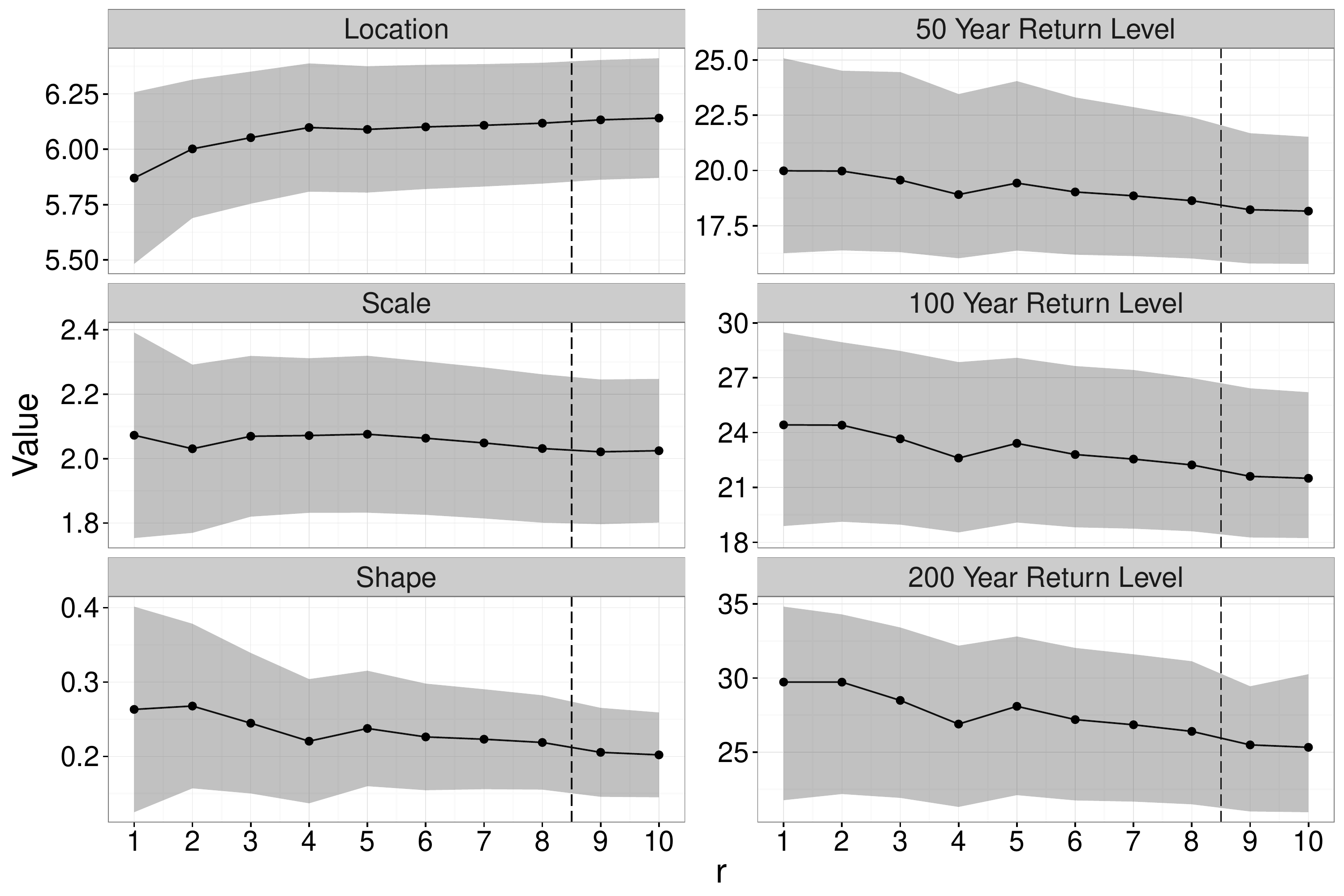}
    \caption{Location, scale, and shape parameter estimates, with 95\% delta 
    confidence intervals for $r=1$ through $r=10$ for the Atlantic City 
    precipitation data. Also included are the estimates and 95\% profile 
    likelihood confidence intervals for the 50, 100, and 200 year return 
    levels.}
    \label{fig:AC_params}
\end{figure*}

With $r=8$, the estimated parameters are given as $\hat{\mu} = 6.118\ (0.139)$, 
$\hat{\sigma}= 2.031\ (0.118)$, and $\hat{\xi}= 0.219\ (0.032)$. This suggests 
a heavy upper tail for the estimated distribution (i.e. $\hat{\xi} > 0$). The 
progression of parameters and certain return level estimates can be seen in 
Figure~\ref{fig:AC_params}. The 50, 100, and 200 year return level 95\% 
confidence intervals for $r=8$ are calculated using the profile likelihood 
method and are given by $(16.019, 22.411)$, $(18.606, 26.979)$, and 
$(21.489, 31.136)$, respectively. The advantages of using $r=8$ versus the 
block maxima for analysis are quite clear from Figure~\ref{fig:AC_params}. 
The standard error of the shape parameter decreases from 0.071 to 0.032, a 
decrease of over 50\%. Similarly, the 50 year return level 95\% 
confidence intervals decreases in width by over 25\%.

\section{Discussion}
\label{s:disc}

We proposed two model specification tests for a fixed number of
largest order statistics as the basis for selecting $r$ for the
$r$ largest order statistics approach in extreme value analysis. 
The score test has two versions of bootstrap procedure: the
multiplier bootstrap method providing a fast, large sample alternative 
to the parametric bootstrap method, with a speedup of over 100 times.
The ED test depends on asymptotic normal approximation of the testing 
statistic, which becomes acceptable for sample size over 50.
It assumes that the $r-1$ top order statistics included already 
fits the GEV$_{r-1}$ distribution. Therefore, the initial 
hypothesis at $r = 1$ needs to be tested with the score tests.
Both tests hold their size better when the shape parameter is 
further away from the lower limit of $-0.5$ or sample size is larger. 
When only small samples are available (50 observations or less), 
the parametric bootstrap score test is recommended.

Alternative versions of the ED test have been explored.
One may define the testing statistics as the difference in entropy 
between GEV$_1$ and GEV$_r$, instead of between GEV$r-1$ and GEV$_r$.
Nonetheless, it appeared to require a larger sample to hold 
its size from our simulation studies (not reported).
In the calculation of $T_n^{(r)}$, the block maxima MLE
$\hat\theta_n^{(1)}$ can be used as an estimate for $\theta$ 
in place of $\hat\theta_n^{(r)}$.
Again, in our simulation studies, this version of the ED test
was too conservative, thus reducing the power,
when the sample size was not large enough.
This may be explained in that the resulting $\hat{S}_{Y_r}$ 
underestimates $S_{Y_r}$.

Naively, the tests may be performed sequentially for each 
$r \in \{1, \ldots, R\}$, for a prefixed, usually small $R$,
at a certain significance level until $H_0^{(r)}$ is rejected.
The issue of multiple, sequential testing is addressed in detail 
by adapting two very recent stopping rules to control the FDR and the 
FWER that are developed specifically for situations when hypotheses 
must be rejected in an ordered fashion \citep{g2015sequential}.
It is shown that these automated procedures reasonably 
control the error rate for the tests discussed in this paper. 
The naive unadjusted sequential testing procedure also 
appears to have decent performance at choosing the correct $r$ 
and for this scenario is more conservative (selects a smaller $r$) 
than the stopping rules.

The tests can be extended to allow covariates in the parameters. 
For example, extremal precipitation in a year may be affected by
large scale climate indexes such as the Southern Oscillation Index 
(SOI), which may be incorporated as a covariate in the location 
parameter \citep[e.g.,][]{Shan:Yan:Zhan:enso:2011}.
Both tests can be carried out with additional model parameters. 
When the underlying data falls into a rich class of dependence 
structures (such as time series), this dependence may be incorporated 
directly instead of using a procedure to achieve approximate 
independence (e.g. the storm length $\tau$ in 
Section~\ref{s:app}). For example, take the GEV-GARCH 
model~\citep{zhao2011garch} when $r=1$. It may be extended
to the case where $r>1$ and the tests presented here may 
be applied to select $r$ under this model assumption.

\appendix

\section{Generating from GEV$_r$ Distribution}
\label{s:gevrsim}

The GEV$_r$ distribution is closely connected to the GEV distribution.
Let $X_1 > \cdots > X_r$ follow a GEV$_r$ distribution~\eqref{eq:gevr}.
It is obvious that the GEV$_1$ distribution is the GEV distribution
with the same parameters, which is the marginal distribution of $X_1$. 
More interestingly, note that, the conditional distribution of $X_2$ 
given $X_1 = x_1$ is simply the GEV distribution righted truncated by $x_1$.
In general, given $(X_1, \ldots, X_k) = (x_1, \ldots, x_k)$ for $1 \le k < r$,
the conditional distribution of $X_{k+1}$ is the GEV distribution
righted truncated at $x_k$.
This property can be exploited to generate the $r$ components 
in a realized GEV$_r$ observation.

The pseudo algorithm to generate a single observation is the following:
\begin{itemize}
  \item
  Generate the first value $x_1$ from the (unconditional) GEV distribution.

  \item
  For $i=2, \ldots, r$:
  
  \begin{itemize}
    \item
    Generate $x_i$ from the GEV distribution right truncated by $x_{i-1}$.
  \end{itemize}
  
\end{itemize}
The resulting vector ($x_1, \ldots, x_r$) is a 
single observation from the GEV$_r$ distribution.

For $\xi \to 0$, caveat is needed in numerical evaluation.
Using function \texttt{expm1} for $\exp(1 + x)$ for $x\to 0$
provides much improved accuracy in comparison to a few 
implementations in existing R packages. 
For readability, here is a simplified version of our implemetation in
R package \texttt{eva} \citep{Rpkg:eva}.
\begin{alltt}
## Quantile function of a GEVr(loc, scale, shape)
qgev <- function(p, loc = 0, scale = 1, shape = 0, 
                 lower.tail = TRUE, log.p = FALSE) \{
  if (log.p) p <- exp(p)
  if(shape == 0) \{
    loc - scale * log(-log(p))
  \} else {
    loc + scale * expm1(log(-log(p)) * -shape) / shape)
  \}
}

## Random number generator of GEVr;
## Returns a matrix of n rows and r columns,
## each row a draw from GEVr
rgevr <- function(n, r, loc = 0, scale = 1, shape = 0) \{
  umat <- matrix(runif(n * r), n, r)
  if (r > 1) \{
    matrix(qgev(t(apply(umat, 1, cumprod)), 
                loc, scale, shape), 
           ncol = r)
  \} else \{
    qgev(umat, loc, scale, shape)
  \}
\}
\end{alltt}

\section{Asymptotic Distribution of $T_{n}^{(r)}(\theta)$}
\label{s:tn}

\begin{proof}[Theorem~\ref{thm:ed}]
Consider a random vector $(X_1, X_2, ... , X_r)$ which follows
a GEV$_r$($\theta$) distribution.
The following result given by \citet[pg. 248]{tawn1988extreme} will be used:
\begin{align}
h(j | \theta, a, b, c) &\equiv E[Z^a_j (1 + \xi Z_j)^{-(\frac{1}{\xi}+b)} \log^c(1 + \xi Z_j)] \notag\\
&=  \frac{(-\xi)^{c-a}}{\Gamma(j)} \sum_{\alpha=0}^{a} (-1)^\alpha {a \choose \alpha} \Gamma^{(c)} (j + b\xi - \alpha \xi + 1)
\label{eq:tawnmoment}
\end{align}
where $Z_j = (X_j - \mu) / \sigma$ and $\Gamma^{(c)}$ is the $c$th 
derivative of the gamma function, for 
$a \in \mathbb{Z}$, $b \in \mathbb{R}$, and $c \in \mathbb{Z}$,
such that $(j + b\xi - \alpha \xi + 1) \not\in \{0, -1, -2, \ldots \}$,
$\alpha = 0, 1, \ldots, a$.

Assume that $\xi \neq 0$ and $1 + \xi Z_j > 0$ for $j=1, \ldots, r$. 
The difference in log-likelihoods for a single observation from the
GEV$_r$($\theta$) and GEV$_{r-1}$($\theta$) distribution is given 
by~\eqref{eq:dll} in Section~\ref{s:ed}. 
Thus, the first moment of $Y_{ir}$ is
\begin{align*}
 E[Y_{1r}] = &\ - \log{\sigma} - h(r | \theta, 0, 0, 0) + h(r - 1 | \theta, 0, 0, 0) \\ 
 &\ - \left(\frac{1}{\xi} + 1\right) h(r | \theta, 0, -\xi^{-1}, 1)  \\
 = &\ -\log{\sigma} - 1 + (1+\xi)\psi(r)
\end{align*}
where $\psi(x) = \frac{\Gamma^{(1)}(x)}{\Gamma(x)}$.

To prove that the second moment of $Y_{ir}$ is finite, note that
\begin{align*}
|Y_{1r}| \leq &\ 4 \max \Bigg\{ \Big|\log{\sigma}\Big|, \Big|(1+\xi Z_{1r})^{-\frac{1}{\xi}}\Big|, \\
&\ \Big|(1+\xi Z_{1{r-1}})^{-\frac{1}{\xi}}\Big|, \Big|\Big(\frac{1}{\xi}+1\Big)\log(1+\xi Z_{1{r-1}})\Big|\Bigg\},
\end{align*}
which implies
\begin{align*}
Y^2_{1r} \leq &\ 16 \Bigg(\max \Bigg\{ \Big|\log{\sigma}\Big|, \Big|(1+\xi Z_{1r})^{-\frac{1}{\xi}}\Big|, \\
&\ \Big|(1+\xi Z_{1{r-1}})^{-\frac{1}{\xi}}\Big|, \Big|\Big(\frac{1}{\xi}+1\Big)\log(1+\xi Z_{1{r-1}})\Big|\Bigg\}\Bigg)^2.
\end{align*}
The bound of $E(Y_{1r}^2)$ can be established by applying~\eqref{eq:tawnmoment} 
to the last three terms in the $\max$ operator,
\begin{align*}
E[(1+\xi Z_{1r})^{-\frac{2}{\xi}}] &= h(r| \theta, 0, \xi^{-1}, 0) < \infty, \\
E[(1+\xi Z_{1{r-1}})^{-\frac{2}{\xi}}] &= h(r-1| \theta, 0, \xi^{-1}, 0) < \infty, \\
E[\log^2(1+\xi Z_{1{r-1}})] &= h(r-1 | \theta, 0, -\xi^{-1}, 2) < \infty.
\end{align*}
The desired result then follows from the central limit theorem and Slutsky's theorem.

The case where $\xi = 0$ in Theorem~\ref{thm:ed} can 
easily be derived by taking the limit as $\xi \to 0$ 
in~\eqref{eq:dll} and in~\eqref{eq:tawnmoment}
by the Dominated Convergence Theorem.
\end{proof}

\begin{acknowledgements}
We would like to thank Dr. Zhiyi Chi for his comments and discussion 
on the multiple testing issues.
\end{acknowledgements}

\bibliographystyle{spbasic}
\bibliography{GEVTwoTests}

\end{document}